\documentclass[12pt]{article}
\usepackage{graphics,epsf,axodraw,amsmath}
\begin{document}

\catcode`\@=11
\@addtoreset{equation}{section}
\@addtoreset{figure}{section}

\newcommand{\qbar}{\overline{q}}
\newcommand{\ubar}{\overline{u}}
\newcommand{\vbar}{\overline{v}}
\newcommand{\lns}[1]{\ln^{#1}\left(\frac{s}{\vec{k}^2}\right)}
\newcommand{\sh}[1]{#1 \!\!\!/}
\newcommand{\Cut}[4]{\SetWidth{0.2}\DashLine(#1,#2)(#3,#4){5}\SetWidth{0.5}}
\newcommand{\hc}{\mathrm{h.c.}}
\renewcommand{\vec}[1]{{\mathbf #1}}

\renewcommand{\theequation}{\thesection.\arabic{equation}}
\renewcommand{\thefigure}{\thesection.\arabic{figure}}

\begin{flushright}
{SHEP-99/08}\\
\end{flushright}
\vspace*{5mm}
\begin{center}
{\large {\bf Studying the Perturbative Reggeon}}\\
\vspace*{1cm}
{\bf S. Griffiths} \\  and \\ {\bf D.A. Ross} \\
\vspace{0.3cm}
Physics Department,\\
University of Southampton,\\
Southampton SO17 1BJ, United Kingdom\\
\vspace*{2cm}
{\bf ABSTRACT} \\ \end{center}
\vspace*{5mm}
\noindent

We consider the flavour non-singlet Reggeon within the context of
perturbative QCD. This consists of ladders built out of ``reggeized''
quarks. We propose a method for the numerical solution of the
integro-differential equation for the amplitude describing the
exchange of such a Reggeon. The solution is known to have a sharp
rise at low values of Bjorken-$x$ when applied to non-singlet
quantities in deep-inelastic scattering.  We show that when the
running of the coupling is taken into account this sharp rise is
further enhanced, although the $Q^2$ dependence is suppressed by the
introduction of the running coupling. We also investigate the effects
of simulating non-perturbative physics by introducing a constituent
mass for the soft quarks and an effective mass for the soft gluons
exchanged in the $t$-channel.

\vspace*{5cm}

\begin{flushleft}
SHEP-99/08 \\
June 1999
\end{flushleft}
\vfill\eject

\setcounter{page}{1}
\pagestyle{plain}

\section{Introduction}

In recent years, much attention has been given to the perturbative QCD
simulation of the Pomeron. The reason for this is that the reach of
HERA is such that one now has data on structure functions and
differential cross-sections for other inclusive processes which are
well into the diffractive region (low-$x$) whilst at the same time
maintaining momentum scales for \emph{all} the kinematic variables.
These variables are large enough that a renormalization-group improved
perturbative expansion summed to all orders in leading $\ln x$ is
expected to be valid.  This is the region in which one expects to be
able to test the BFKL Pomeron \cite{bfkl}.  Notwithstanding this, it
should be recalled that the original motivation for the study of the
Pomeron in QCD was an attempt to explain how the successes of Regge
theory could be under-written by a quantum field theory.

Somewhat less emphasis has been placed on the Regge trajectory below
the Pomeron (the Reggeon) for which phenomenological fits
\cite{landdonn} have shown have an intercept $\approx \frac{1}{2}$.
In the same way that the QCD Pomeron is constructed from ladders of
reggeized gluons with a colour singlet projection, so the QCD Reggeon
is constructed from ladders of a reggeized quark-antiquark pair, again
with colour singlet projection.

The cleanest way to distinguish between Pomeron dominated processes
and Reggeon dominated processes is to note that a quark-antiquark pair
can be in a flavour non-singlet state. We therefore consider
quantities which are controlled by flavour non-singlet operators. Such
quantities include the structure function $F_3$ in deep-inelastic
scattering, or the spin-dependent structure functions.  The Reggeon is
only expected to dominate such quantities at sufficiently low-$x$, and
the extraction of low-$x$ data for such quantities is very difficult.
The spin-dependent structure functions at low-$x$ have recently been
considered in ref.\cite{kwiecinski}. At the moment the lowest values
of $x$ for which we have data on $F_3$ are not yet in the asymptotic
region where we would expect the Reggeon to dominate, but on the other
hand they are not very far away and in the same way that HERA is
currently used to test the QCD Pomeron, the QCD Reggeon could
reasonably be expected to be probed in the not-too-distant future.

In order to construct the Reggeon within the context of perturbative
QCD, one needs first to establish that the quark reggeizes, in the
same way that the gluon reggeizes. In other words, it is necessary to
show that to leading order in $\ln s$, the amplitude in which the
quantum numbers of a quark is exchanged in the $t$-channel, has an $s$
dependence given by
$$s^{\alpha_\mathcal{Q}(t)}.$$
From standard partial wave analysis we expect
$\alpha_\mathcal{Q}(t)=\frac{1}{2}$ in leading order and for the $t$
dependence to arise from higher order corrections. This was first
established some time ago \cite{fadinsherman}. However, we feel that
several important features of this derivation were not sufficiently
clearly explained in ref.\cite{fadinsherman}. We have repeated the
relevant calculations and we briefly review our method for deriving
the reggeization of the quark in section 2. Once quark reggeization
has been established, an integro-differential equation can be
established for the sum of the leading $\ln s$ parts of the amplitude
for the exchange of the quantum numbers of the Reggeon. This has been
derived in ref.\cite{kirschner} and we quote the result in section 3.
This equation can, in principle, be solved analytically since it
involves a kernel which is conformally invariant (in two dimensions)
such that its eigenfunctions are the representations of the conformal
group. Unlike the case of the Pomeron the leading eigenvalue is
\emph{not} analytic in the coupling. This causes severe difficulties
when one tries to reproduce the analytic results by numerical means.
We propose a programme for numerical solution of the
integro-differential equation which is consistent with the known
asymptotic behaviour. This allows one to introduce modifications to
the kernel, such as the running of the coupling, which breaks
conformal invariance, so that the analytic approach is no longer
viable, but the numerical approach remains reliable.

In section 4 we apply this to a model of flavour non-singlet
deep-inelastic scattering.  We call this a model since an arbitrary
function has to be taken for the impact factor describing the emission
of a Reggeon from the target hadron. However, we do not expect either
the $x$ dependence at sufficiently low $x$, or the $Q^2$ at
sufficiently high $Q^2$ to be particularly sensitive to the exact
nature of this impact factor. We look at the effect of running the
coupling and observe that whereas this further enhances the predicted
rise in the non-singlet structure functions as $ x \to 0$, it
suppresses the $Q^2$ dependence of such structure functions.  Finally,
we postulate that the most significant feature of non-perturbative
effects in the case of the Reggeon is to provide the soft quark and
antiquark exchanged in the $t$-channel with a constituent mass, and
the soft gluons an effective mass. We therefore solve the modified
equation for the Reggeon in which a mass is assigned to the soft
particles. We find that this significantly suppresses the low-$x$ rise
and that this suppression extends to values of $Q^2$ way beyond the
assigned values of the constituent mass.  In section 5 we summarize
our conclusions.

                  
\section{The Reggeized Quark}

In this section we review the derivation of the ``reggeized quark''. 
The result obtained agrees with that previously obtained \cite{fadinsherman},
but in our derivation some of the subtleties leading to the result 
are discussed in more detail.
\begin{figure}
\begin{center}
  \begin{picture}(200,100)(-20,0)
    \ArrowLine(20,80)(80,80)
    \Vertex(80,80){1.4}
    \ArrowLine(80,80)(80,20)
    \Vertex(80,20){1.4}
    \ArrowLine(80,20)(20,20) 
    \Gluon(80,80)(140,80){3}{6}
    \Text(150,80)[l]{$\epsilon_1$}    
    \Gluon(80,20)(140,20){3}{6}
    \Text(150,20)[l]{$\epsilon_2$}    
    \LongArrow(10,85)(25,85)
    \Text(5,85)[r]{$p_1$}
    \LongArrow(10,15)(25,15)
    \Text(5,15)[r]{$p_2$}
    \LongArrow(75,58)(75,43)
    \Text(70,50)[r]{$q$}
    \Text(110,70)[c]{$a$}
    \Text(110,30)[c]{$b$}
    \Text(20,70)[c]{$j$}
    \Text(20,30)[c]{$i$}
  \end{picture}
\end{center}
  \caption{The tree-level exchange of a soft quark}
  \label{fig1}
\end{figure}
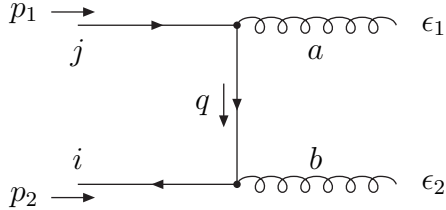

We begin by considering the process
\begin{equation}
  q \ + \ \bar{q} \ \to \ g \ + \ g, \label{process1}
\end{equation} 
in the Regge regime, $s\gg|t|$, where a soft quark is exchanged in
the $t$-channel as shown in Fig.~\ref{fig1}.
The amplitude is given by 
\begin{equation}
  {\cal A}_0(q)=(\tau^b\tau^a)_{ij} 2\pi\alpha_s\vbar(p_2)\gamma
  u(p_1)\frac{\epsilon_2^*q\epsilon_1^*}{|q|^2} + \hc,
  \label{eq:tree-level2}
\end{equation}
where we have introduced holomorphic coordinates in the plane
transverse to the incident momenta $p_1$ and $p_2$ such that $\gamma
\, (\gamma^*)$ are the $\gamma$-matrices in this plane, $q \,(q^*)$
are the components of the transverse momentum transferred in the
$t$-channel, and $\epsilon_1 \,(\epsilon_1^*)$ and $\epsilon_2
\,(\epsilon_2^*)$ are the polarisation vectors of the outgoing gluons.

We now consider the one-loop corrections which are proportional to
$\ln s$, since it is only these contributions that are relevant for
reggeization.  The relevant diagrams are shown in Fig.~\ref{fig3}, and
each of these gives the same contribution (up to a colour factor),
which may be written as \footnote{ The quantity $\vec{k}^2$, used to
  scale $s$ inside the logarithms is understood to be a square
  momentum which is of the order of the square momentum transfer $|t|$,
  or some other momentum ($\ll s$) involved in the coupling of the
  Reggeon at the top or bottom of the ladder. Since we are confining
  our discussion to leading logarithm, its exact value is
  unimportant.}
\begin{equation}
  \label{eq:one-loop} 
  \frac{\alpha_s^2}{\pi}\lns{}\vbar(p_2)\gamma
  u(p_1)\epsilon_2^*\epsilon_1^*
  \int d^2\vec{k}\frac{k}{|k|^2|q-k|^2} +\hc
\end{equation} 
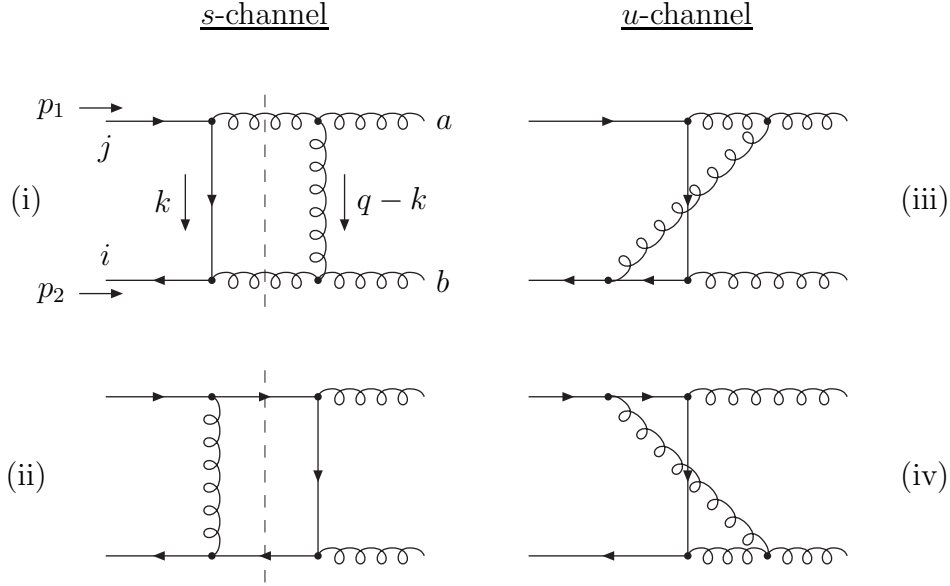
\begin{figure}
  \begin{center}
    \[
    \begin{array}{cc}
    \begin{picture}(150,140)(0,0)
      \ArrowLine(20,80)(60,80)
      \Vertex(60,80){1.4}
      \ArrowLine(60,80)(60,20)
      \Vertex(60,20){1.4}
      \ArrowLine(60,20)(20,20)
      \Gluon(60,80)(100,80){3}{4}
      \Gluon(100,80)(140,80){3}{4}
      \Gluon(60,20)(100,20){3}{4}
      \Gluon(100,20)(140,20){3}{4}
      \Gluon(100,80)(100,20){3}{6}
      \Vertex(100,80){1.4}
      \Vertex(100,20){1.4}
      \LongArrow(10,85)(25,85)
      \Text(5,85)[r]{$p_1$}
      \LongArrow(10,15)(25,15)
      \Text(5,15)[r]{$p_2$}
      \LongArrow(50,60)(50,40)
      \Text(45,50)[r]{$k$}    
      \LongArrow(110,60)(110,40)
      \Text(115,50)[l]{$q-k$}  
      \Text(145,80)[l]{$a$}
      \Text(145,20)[l]{$b$}
      \Text(20,70)[c]{$j$}
      \Text(20,30)[c]{$i$}
      \Text(-10,50)[c]{(i)}
      \Text(80,120)[c]{\underline{$s$-channel}}
      \Cut{80}{10}{80}{90}
    \end{picture}    
&
    \begin{picture}(150,140)(0,0)
      \ArrowLine(20,80)(80,80)
      \Vertex(80,80){1.4}
      \ArrowLine(80,80)(80,20)
      \Vertex(80,20){1.4}
      \ArrowLine(80,20)(50,20)
      \ArrowLine(50,20)(20,20)
      \Gluon(80,80)(110,80){3}{3}
      \Vertex(110,80){1.4}
      \Gluon(110,80)(140,80){3}{3}
      \Gluon(80,20)(140,20){3}{6}
      \Gluon(110,80)(50,20){3}{8}
      \Vertex(50,20){1.4}
      \Text(170,50)[c]{(iii)}
      \Text(80,120)[c]{\underline{$u$-channel}}
    \end{picture}
\\
    \begin{picture}(150,100)(0,0)
      \Vertex(60,80){1.4}
      \Vertex(100,80){1.4}
      \Vertex(100,20){1.4}
      \Vertex(60,20){1.4}
      \ArrowLine(20,80)(60,80)
      \ArrowLine(60,80)(100,80)
      \ArrowLine(100,80)(100,20)
      \ArrowLine(100,20)(60,20)
      \ArrowLine(60,20)(20,20)
      \Gluon(100,80)(140,80){3}{4}
      \Gluon(100,20)(140,20){3}{4}
      \Gluon(60,80)(60,20){3}{6}
      \Text(-10,50)[c]{(ii)}
      \Cut{80}{10}{80}{90}
    \end{picture}
&
    \begin{picture}(150,100)(0,0)
      \ArrowLine(20,80)(50,80)
      \Vertex(50,80){1.4}
      \ArrowLine(50,80)(80,80)
      \Vertex(80,80){1.4}
      \ArrowLine(80,80)(80,20)
      \Vertex(80,20){1.4}
      \ArrowLine(80,20)(20,20)
      \Gluon(80,20)(110,20){3}{3}
      \Vertex(110,20){1.4}
      \Gluon(110,20)(140,20){3}{3}
      \Gluon(80,80)(140,80){3}{6}
      \Gluon(50,80)(110,20){3}{8}
      \Text(170,50)[c]{(iv)}
    \end{picture}
    \end{array}
    \]
  \end{center}
  \caption{One-loop leading $\ln s$ corrections to $q\ + \qbar\
    \rightarrow\ g\ +\ g$. The 2-body cuts
    in the $s$-channel are shown, however (iii) and (iv) have
    similar cuts in the $u$-channel}
  \label{fig3}
\end{figure}
The diagrams of Fig.~\ref{fig3} have colour factors
\begin{eqnarray}  
\frac{N}{2} \left( \tau^b \tau^a \right)_{ij} + \frac{1}{4} \delta^{ab} \delta_{ij}
 \  & \ \mathrm{(i)} \nonumber  \\  
\frac{-1}{2N} \left( \tau^b \tau^a \right)_{ij} + \frac{1}{4} \delta^{ab} \delta_{ij}
 \  & \ \mathrm{(ii)} \nonumber  \\  
 -  \ \frac{1}{4} \delta^{ab} \delta_{ij}
 \  & \ \mathrm{(iii)} \nonumber  \\  
 -  \ \frac{1}{4} \delta^{ab} \delta_{ij}
 \  & \ \mathrm{(iv)} \nonumber  \end{eqnarray}  
where $N \ (=3)$ is the number of colours.

Adding these one-loop contributions we obtain the result for the
one-loop correction to be
\begin{equation}
  \label{eq:one-loop-result}
  {\cal A}_1(q)={\cal A}_0(q)\lns{}\epsilon_\mathcal{Q}(q)
\end{equation}
with
\begin{equation}
  \label{eq:epsilon}
  \epsilon_\mathcal{Q}(q)=-C_F\frac{\alpha_s}{2\pi^2}\int 
  d^2\vec{k}\frac{kq^*}{|k|^2|q-k|^2},
\end{equation}
and $C_F=(N^2-1)/2N$.

\begin{figure}
  \[
  \begin{array}{c}
  \begin{array}{ccc}
    \begin{picture}(140,100)(0,0)
      \Vertex(80,20){1.4}
      \Vertex(80,80){1.4}
      \ArrowLine(20,80)(80,80)
      \LongArrow(10,85)(25,85)
      \Text(5,85)[r]{$p_1$}
      \ArrowLine(80,80)(80,50)
      \LongArrow(70,72)(70,58)
      \Text(60,65)[r]{$k_{i-1}$}
      \ArrowLine(80,50)(80,20)
      \LongArrow(70,42)(70,28)
      \Text(60,35)[r]{$k_{i}$}
      \ArrowLine(80,20)(20,20)
      \LongArrow(10,15)(25,15)
      \Text(5,15)[r]{$p_2$}
      \Gluon(80,80)(140,80){3}{5}
      \Gluon(80,20)(140,20){3}{5}
      \Gluon(80,50)(140,50){3}{5}
      \GOval(80,50)(5,5)(0){0.4}
      \Text(145,80)[l]{$\mu$}
      \Text(145,50)[l]{$\sigma$}
      \Text(145,20)[l]{$\nu$}
      \Text(110,90)[c]{$a$}
      \Text(110,60)[c]{$c$}
      \Text(110,30)[c]{$b$}
    \end{picture}
&
    \begin{picture}(20,20)(-20,-50)
      \Text(0,0)[c]{$=$}
    \end{picture}
&
    \SetScale{0.5}
   \begin{picture}(60,50)(0,-25)
      \Vertex(80,20){1.4}
      \Vertex(80,80){1.4}
      \Vertex(80,50){1.4}
      \ArrowLine(20,80)(80,80)
      \ArrowLine(80,80)(80,50)
      \ArrowLine(80,50)(80,20)
      \ArrowLine(80,20)(20,20)
      \Gluon(80,80)(140,80){5}{3}
      \Gluon(80,20)(140,20){5}{3}
      \Gluon(80,50)(140,50){5}{3}
    \end{picture}
    \end{array}
\\
    \begin{array}{cccccccc}
    \begin{picture}(20,20)(-20,-25)
      \Text(0,0)[c]{$+$}
    \end{picture}
&
    \SetScale{0.5}
    \begin{picture}(60,50)(0,0)
      \Vertex(80,20){1.4}
      \Vertex(80,80){1.4}
      \Vertex(110,20){1.4}
      \ArrowLine(20,80)(80,80)
      \ArrowLine(80,80)(80,20)
      \ArrowLine(80,20)(20,20)
      \Gluon(80,80)(140,80){5}{3}
      \Gluon(80,20)(110,20){5}{1}
      \Gluon(110,20)(140,20){5}{1}
      \GlueArc(140,20)(30,90,180){5}{2}
    \end{picture}
&    
    \begin{picture}(20,20)(-20,-25)
      \Text(0,0)[c]{$+$}
    \end{picture}
&
    \SetScale{0.5}
    \begin{picture}(60,50)(0,0)
      \Vertex(80,20){1.4}
      \Vertex(80,80){1.4}
      \Vertex(50,20){1.4}
      \Gluon(80,80)(140,80){5}{3}
      \ArrowLine(80,80)(80,20)
      \ArrowLine(20,80)(80,80)
      \Gluon(80,20)(140,20){5}{3}
      \ArrowLine(50,20)(20,20)
      \ArrowLine(80,20)(50,20)
      \Gluon(50,20)(140,50){5}{4}
    \end{picture}
&
    \begin{picture}(20,20)(-20,-25)
      \Text(0,0)[c]{$+$}
    \end{picture}
&
    \SetScale{0.5}
    \begin{picture}(60,50)(0,0)
      \Vertex(80,20){1.4}
      \Vertex(80,80){1.4}
      \Vertex(50,80){1.4}
      \Gluon(80,20)(140,20){5}{3}
      \ArrowLine(80,80)(80,20)
      \ArrowLine(80,20)(20,20)
      \Gluon(80,80)(140,80){5}{3}
      \ArrowLine(20,80)(50,80)
      \ArrowLine(50,80)(80,80)
      \Gluon(140,50)(50,80){5}{4}
    \end{picture}
&
    \begin{picture}(20,20)(-20,-25)
      \Text(0,0)[c]{$+$}
    \end{picture}
&
    \SetScale{0.5}
    \begin{picture}(60,50)(0,0)
      \Vertex(80,20){1.4}
      \Vertex(80,80){1.4}
      \ArrowLine(80,20)(20,20)
      \ArrowLine(80,80)(80,20)
      \ArrowLine(20,80)(80,80)
      \Gluon(80,20)(140,20){5}{3}
      \Gluon(80,80)(110,80){5}{1}
      \Vertex(110,80){1.4}
      \Gluon(110,80)(140,80){5}{1}
      \GlueArc(140,80)(30,180,270){5}{2}
    \end{picture}
  \end{array}
  \end{array}
  \]
  \caption{The effective quark-gluon vertex}
  \label{fig4}
\end{figure}
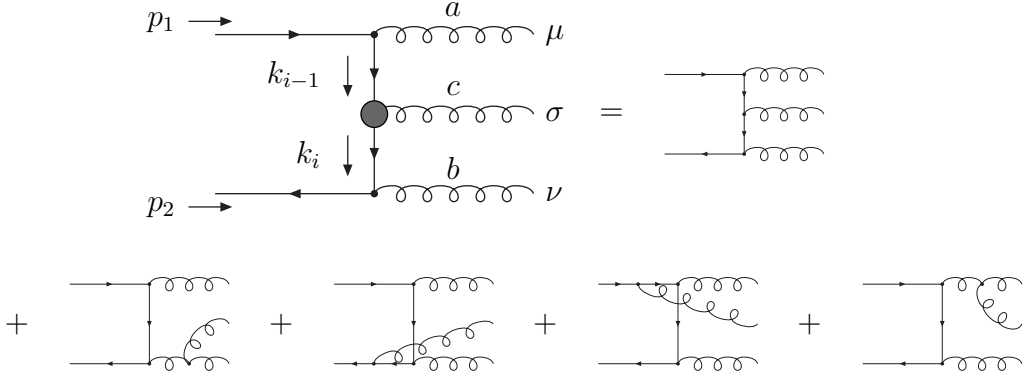

At the next order a number of complications occur, some of which are
common to the problem of gluon reggeization \cite{bfkl}. The first of
these is that the diagrams that contribute in leading $\ln s$ are {\it not}
all of the ladder type. However, all diagrams whose $s$-channel cuts
 contain  three intermediate particles can be reduced to a ladder-type 
diagram using an effective vertex. This is demonstrated in Fig.~\ref{fig4}.
The Feynman rule for this effective vertex involves the substitution
of $\gamma^\sigma$ in the standard rule with $\Gamma_{\mathcal
  Q}^\sigma$, where
\begin{equation}
  \Gamma_{\mathcal Q}^\sigma(k_{i-1},k_i)=\gamma^\sigma +
  \frac{2p_1^\sigma\sh{k}_{i-1}}{\beta_{i}s}
  +\frac{2p_2^\sigma\sh{k}_{i}}{\alpha_{i-1}s}   
\end{equation}
from which it follows that
\begin{equation}
  \vbar(p_2)\left(\Gamma_{\mathcal{Q}}\cdot\epsilon_i\right)u(p_1) 
  =\vbar(p_2)\gamma u(p_1)\frac{1}{2}
  \left(\frac{\epsilon_i k_{i-1}^*}{k_{i-1}-k_i} +
    \frac{\epsilon_i^* k_i^*}{k_{i-1}^*-k_i^*}
  \right) + \hc \label{quarkeffective}
\end{equation}
$k_{i-1}$, $k_i$ are the transverse momenta of the incoming and outgoing
quarks, defined in terms of Sudakov variables by
\begin{equation}
 k_{i}^\mu=\alpha_{i}\, p_1^\mu + \beta_{i}\, p_2^\mu
 +k_{\perp i}^\mu \nonumber
 \end{equation}
with $k_{\perp i}^2=-|k_i|^2$.

We also need the equivalent effective vertex for the emission of a
gluon with polarisation $\epsilon_i$ from incoming and outgoing
vertical gluon lines with momenta $k_{i-1}$, $k_i$ and Lorentz indices
$\mu$, $\nu$. This is given by
\begin{equation}
  \Gamma_{\mathcal{G}}^{\mu\nu}(k_i,k_{i-1})\cdot{\epsilon}_i=
  \frac{2p_2^\mu p_1^\nu}{s}\left(
    \frac{\epsilon_i k_{i-1}^*k_i}{k_i-k_{i-1}}+\hc \right).
\end{equation}
In both of these effective vertices, we have assumed the multi-Regge
kinematics, namely
$$ \alpha_{i-1} \ \gg  \ \alpha_i $$
$$ |\beta_i| \ \gg \ |\beta_{i-1}|. $$

The product of these two effective vertices, summed over the polarisations
for the emitted gluon is given by
\begin{equation}
  \sum_{\rm{pol'ns}}(\Gamma_{\mathcal{G}}\cdot\epsilon)
  (\Gamma_{\mathcal{Q}}\cdot\epsilon)=
  \gamma\left(q^*-\frac{{k'}^*|q-k|^2}{|k-k'|^2}
  -\frac{k^*|q-k'|^2}{|k-k'|^2}\right) +\hc
\end{equation}

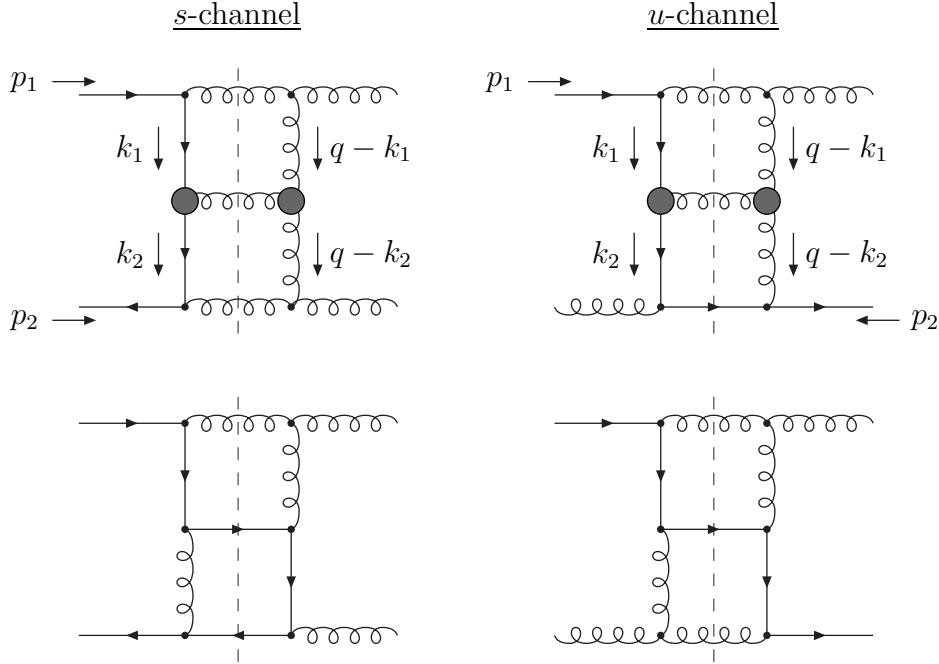
\begin{figure}
  \leavevmode
  \SetScale{1}
  \[
  \begin{array}{cc}
    \begin{picture}(170,120)(0,0)
      \label{fig:two-loop-main2}
      \ArrowLine(20,90)(60,90)
      \Vertex(60,90){1.4}
      \ArrowLine(60,90)(60,50)
      \ArrowLine(60,50)(60,10)
      \Vertex(60,10){1.4}
      \ArrowLine(60,10)(20,10)
      \Gluon(60,90)(100,90){3}{4}
      \Vertex(100,90){1.4}
      \Gluon(100,90)(140,90){3}{4}
      \Gluon(60,10)(100,10){3}{4}
      \Vertex(100,10){1.4}
      \Gluon(100,10)(140,10){3}{4}
      \Gluon(100,90)(100,50){3}{3}
      \Gluon(100,50)(100,10){3}{3}
      \Gluon(60,50)(100,50){3}{4}
      \GOval(60,50)(5,5)(0){0.4} \GOval(100,50)(5,5)(0){0.4}
      \LongArrow(10,95)(25,95)
      \Text(5,95)[r]{$p_1$}
      \LongArrow(10,5)(25,5)
      \Text(5,5)[r]{$p_2$}
      \LongArrow(50,38)(50,23)
      \Text(45,30)[r]{$k_2$}
      \LongArrow(110,78)(110,63)
      \Text(115,70)[l]{$q-k_1$}
      \LongArrow(50,78)(50,63)
      \Text(45,70)[r]{$k_1$}
      \LongArrow(110,38)(110,23)
      \Text(115,30)[l]{$q-k_2$}
      \Cut{80}{100}{80}{0}
      \Text(80,120)[c]{\underline{$s$-channel}}
    \end{picture}   
&
     \begin{picture}(170,120)(0,0)
      \ArrowLine(20,90)(60,90)
      \Vertex(60,90){1.4}
      \ArrowLine(60,90)(60,50)
      \ArrowLine(60,50)(60,10)
      \Vertex(60,10){1.4}
      \ArrowLine(60,10)(100,10)
      \ArrowLine(100,10)(140,10)
      \Gluon(60,90)(100,90){3}{4}
      \Vertex(100,90){1.4}
      \Gluon(100,90)(140,90){3}{4}
      \Gluon(60,10)(20,10){3}{4}
      \Vertex(100,10){1.4}
      \Gluon(100,90)(100,50){3}{3}
      \Gluon(100,50)(100,10){3}{3}
      \Gluon(60,50)(100,50){3}{4}
      \GOval(60,50)(5,5)(0){0.4} \GOval(100,50)(5,5)(0){0.4}
      \LongArrow(10,95)(25,95)
      \Text(5,95)[r]{$p_1$}
      \LongArrow(150,5)(135,5)
      \Text(155,5)[l]{$p_2$}
      \LongArrow(50,38)(50,23)
      \Text(45,30)[r]{$k_2$}
      \LongArrow(110,78)(110,63)
      \Text(115,70)[l]{$q-k_1$}
      \LongArrow(50,78)(50,63)
      \Text(45,70)[r]{$k_1$}
      \LongArrow(110,38)(110,23)
      \Text(115,30)[l]{$q-k_2$}
      \Cut{80}{100}{80}{0}
      \Text(80,120)[c]{\underline{$u$-channel}}
    \end{picture}
\\
    \begin{picture}(170,120)(0,0)
      \ArrowLine(20,90)(60,90)
      \ArrowLine(100,10)(60,10)
      \ArrowLine(60,90)(60,50)
      \ArrowLine(60,50)(100,50)
      \ArrowLine(100,50)(100,10)
      \ArrowLine(60,10)(20,10)
      \Vertex(60,90){1.4}
      \Gluon(100,90)(100,50){3}{3}
      \Vertex(60,10){1.4}
      \Vertex(100,90){1.4}
      \Gluon(100,90)(140,90){3}{4}
      \Gluon(60,90)(100,90){3}{4}
      \Vertex(100,10){1.4}
      \Gluon(100,10)(140,10){3}{4}
      \Gluon(60,50)(60,10){3}{3}
      \Vertex(60,50){1.4}
      \Vertex(100,50){1.4}
      \Cut{80}{100}{80}{0}
    \end{picture}
&
    \begin{picture}(170,120)(0,0)
      \ArrowLine(20,90)(60,90)
      \Gluon(100,10)(60,10){3}{4}
      \ArrowLine(60,90)(60,50)
      \ArrowLine(60,50)(100,50)
      \ArrowLine(100,50)(100,10)
      \Gluon(60,10)(20,10){3}{4}
      \Vertex(60,90){1.4}
      \Gluon(100,90)(100,50){3}{3}
      \Vertex(60,10){1.4}
      \Vertex(100,90){1.4}
      \Gluon(100,90)(140,90){3}{4}
      \Gluon(60,90)(100,90){3}{4}
      \Vertex(100,10){1.4}
      \ArrowLine(100,10)(140,10)
      \Gluon(60,50)(60,10){3}{3}
      \Vertex(60,50){1.4}
      \Vertex(100,50){1.4}
      \Cut{80}{100}{80}{0}
    \end{picture}
  \end{array}
  \]
  \caption{Examples of 2-loop ladders. Diagrams with a quark or
    antiquark rung have to be considered in addition to the gluon
    rungs with effective vertices} 
  \label{fig5}
\end{figure}

A further complication arises from the need to consider diagrams in
which the central cut line involves a quark or antiquark (as opposed
to a gluon) and this is demonstrated pictorially in Fig.~\ref{fig5}.

\begin{figure}
  \SetScale{0.60}
  \begin{center}
    \leavevmode
    \[
    \begin{array}{cccc}
      \SetScale{1.0}
      \begin{picture}(50,100)(0,0)
       \ArrowLine(20,80)(40,80)
        \Vertex(40,80){1.4}
        \ArrowLine(40,80)(40,20)
        \Vertex(40,20){1.4}
        \ArrowLine(40,20)(20,20)
        \Gluon(40,80)(80,80){3}{3}
        \Vertex(80,80){1.4}
        \Gluon(80,80)(120,80){3}{3}
        \Vertex(120,80){1.4}
        \Gluon(120,80)(140,80){3}{2}
        \Gluon(40,20)(80,20){3}{3}
        \Vertex(80,20){1.4}
        \Gluon(80,20)(120,20){3}{3}
        \Vertex(120,20){1.4}
        \Gluon(120,20)(140,20){3}{2}  
        \Gluon(80,80)(80,20){3}{6}
        \Gluon(120,80)(120,20){3}{6}
        \LongArrow(10,85)(25,85)
        \Text(5,85)[r]{$p_1$}
        \LongArrow(10,15)(25,15)
        \Text(5,15)[r]{$p_2$}
        \LongArrow(30,58)(30,43)
        \Text(25,50)[r]{$k_2$}
        \LongArrow(130,58)(130,43)
        \Text(135,50)[l]{\rotatebox{90}{$q-k_1$}}
        \LongArrow(70,58)(70,43)
        \Text(65,50)[r]{\rotatebox{90}{$k_1-k_2$}}
        \Cut{50}{90}{50}{10}
        \Cut{100}{90}{100}{10}
        \Text(80,120)[c]{\underline{$s$-channel}}
      \end{picture}
&&
      \SetScale{1.0}
      \begin{picture}(50,100)(0,0)
        \ArrowLine(20,80)(40,80)
        \Vertex(40,80){1.4}
        \ArrowLine(40,80)(40,20)
        \Vertex(40,20){1.4}
        \Gluon(40,20)(20,20){3}{2}
        \Gluon(40,80)(80,80){3}{3}
        \Vertex(80,80){1.4}
        \Gluon(80,80)(120,80){3}{3}
        \Vertex(120,80){1.4}
        \Gluon(120,80)(140,80){3}{2}
        \ArrowLine(40,20)(80,20)
        \Vertex(80,20){1.4}
        \ArrowLine(80,20)(120,20)
        \Vertex(120,20){1.4}
        \ArrowLine(120,20)(140,20)  
        \Gluon(80,80)(80,20){3}{6}
        \Gluon(120,80)(120,20){3}{6}
        \LongArrow(10,85)(25,85)
        \Text(5,85)[r]{$p_1$}
        \LongArrow(150,15)(135,15)
        \Text(155,15)[l]{$p_2$}
        \LongArrow(30,58)(30,43)
        \Text(25,50)[r]{$k_2$}
        \LongArrow(130,58)(130,43)
        \Text(135,50)[l]{\rotatebox{90}{$q-k_1$}}
        \LongArrow(70,58)(70,43)
        \Text(65,50)[r]{\rotatebox{90}{$k_1-k_2$}}
        \Cut{50}{90}{50}{10}
        \Cut{100}{90}{100}{10}
        \Text(80,120)[c]{\underline{$u$-channel}}
      \end{picture}
&\\
      \begin{picture}(80,80)(0,0)
        \Vertex(30,80){1.4}
        \Vertex(60,80){1.4}
        \Vertex(80,80){1.4}
        \Vertex(30,20){1.4}
        \Vertex(60,20){1.4}
        \Vertex(80,20){1.4}
        \Line(20,80)(30,80)
        \ArrowLine(30,80)(60,80)
        \ArrowLine(60,80)(80,80)
        \ArrowLine(80,80)(80,20)
        \ArrowLine(80,20)(60,20)
        \ArrowLine(60,20)(30,20)
        \Line(30,20)(20,20)
        \Gluon(80,80)(140,80){3}{6}
        \Gluon(80,20)(140,20){3}{6}
        \Gluon(30,80)(30,20){3}{6}
        \Gluon(60,80)(60,20){3}{6}
        \Cut{45}{90}{45}{10}
        \Cut{70}{90}{70}{10}
      \end{picture}
&
      \begin{picture}(100,80)(0,0)
        \Vertex(50,80){1.4}
        \Vertex(80,80){1.4}
        \Vertex(50,20){1.4}
        \Vertex(80,20){1.4}
        \Vertex(110,80){1.4}
        \Vertex(110,20){1.4}
        \ArrowLine(20,80)(50,80)
        \ArrowLine(50,80)(80,80)
        \ArrowLine(50,20)(20,20)
        \ArrowLine(80,80)(80,20)
        \ArrowLine(80,20)(50,20)
        \Gluon(80,20)(110,20){3}{3}
        \Gluon(110,20)(140,20){3}{3}
        \Gluon(80,80)(110,80){3}{3}
        \Gluon(110,80)(140,80){3}{3}
        \Gluon(50,80)(50,20){3}{6}
        \Gluon(110,20)(110,80){3}{8}
        \Cut{65}{90}{65}{10}
        \Cut{95}{90}{95}{10}  
      \end{picture}
&
      \begin{picture}(80,80)(0,0)
        \Vertex(30,80){1.4}
        \Vertex(60,80){1.4}
        \Vertex(80,80){1.4}
        \Vertex(30,20){1.4}
        \Vertex(60,20){1.4}
        \Vertex(80,20){1.4}
        \Line(20,80)(30,80)
        \ArrowLine(30,80)(60,80)
        \ArrowLine(60,80)(80,80)
        \ArrowLine(80,80)(80,20)
        \Gluon(80,20)(60,20){3}{2}
        \Gluon(60,20)(30,20){3}{2}
        \Gluon(30,20)(20,20){3}{1}
        \Gluon(80,80)(140,80){3}{6}
        \ArrowLine(80,20)(140,20)
        \Gluon(30,80)(30,20){3}{6}
        \Gluon(60,80)(60,20){3}{6}
        \Cut{45}{90}{45}{10}
        \Cut{70}{90}{70}{10}
      \end{picture}
&
      \begin{picture}(80,80)(0,0)
        \Vertex(50,80){1.4}
        \Vertex(80,80){1.4}
        \Vertex(50,20){1.4}
        \Vertex(80,20){1.4}
        \Vertex(110,80){1.4}
        \Vertex(110,20){1.4}
        \ArrowLine(20,80)(50,80)
        \ArrowLine(50,80)(80,80)
        \Gluon(50,20)(20,20){3}{3}
        \ArrowLine(80,80)(80,20)
        \Gluon(80,20)(50,20){3}{3}
        \ArrowLine(80,20)(110,20)
        \ArrowLine(110,20)(140,20)
        \Gluon(80,80)(110,80){3}{3}
        \Gluon(110,80)(140,80){3}{3}
        \Gluon(50,80)(50,20){3}{6}
        \Gluon(110,20)(110,80){3}{8}
        \Cut{65}{90}{65}{10}
        \Cut{95}{90}{95}{10}  
      \end{picture}
\\
      \begin{picture}(80,80)(0,0)
        \Vertex(80,80){1.4}
        \Vertex(80,20){1.4}
        \Vertex(100,80){1.4}
        \Vertex(120,80){1.4}
        \Vertex(100,20){1.4}
        \Vertex(120,20){1.4}
        \ArrowLine(20,80)(80,80)
        \ArrowLine(80,80)(80,20)
        \ArrowLine(80,20)(20,20)
        \Gluon(80,80)(100,80){3}{2}
        \Gluon(100,80)(120,80){3}{2}
        \Gluon(120,80)(140,80){3}{2}
        \Gluon(80,20)(100,20){3}{2}
        \Gluon(100,20)(120,20){3}{2}
        \Gluon(120,20)(140,20){3}{2}
        \Gluon(100,80)(120,20){3}{6}
        \Gluon(120,80)(100,20){3}{6}
        \Cut{90}{90}{90}{10}
      \end{picture}
&
      \begin{picture}(100,80)(0,0)
        \Vertex(80,80){1.4}
        \Vertex(80,20){1.4}
        \Vertex(30,80){1.4}
        \Vertex(30,20){1.4}
        \Vertex(60,80){1.4}
        \Vertex(60,20){1.4}
        \Line(20,80)(30,80)
        \ArrowLine(30,80)(60,80)
        \ArrowLine(60,80)(80,80)
        \ArrowLine(80,80)(80,20)
        \ArrowLine(80,20)(60,20)
        \ArrowLine(60,20)(30,20)
        \Line(30,20)(20,20)
        \Gluon(80,80)(140,80){3}{6}
        \Gluon(80,20)(140,20){3}{6}
        \Gluon(30,80)(60,20){3}{6}
        \Gluon(60,80)(30,20){3}{6}
        \Cut{70}{90}{70}{10}
      \end{picture}
&
      \begin{picture}(80,80)(0,0)
        \Vertex(80,80){1.4}
        \Vertex(80,20){1.4}
        \Vertex(100,80){1.4}
        \Vertex(120,80){1.4}
        \Vertex(100,20){1.4}
        \Vertex(120,20){1.4}
        \ArrowLine(20,80)(80,80)
        \ArrowLine(80,80)(80,20)
        \Gluon(80,20)(20,20){3}{6}
        \Gluon(80,80)(100,80){3}{2}
        \Gluon(100,80)(120,80){3}{2}
        \Gluon(120,80)(140,80){3}{2}
        \ArrowLine(80,20)(100,20)
        \ArrowLine(100,20)(120,20)
        \ArrowLine(120,20)(140,20)
        \Gluon(100,80)(120,20){3}{6}
        \Gluon(120,80)(100,20){3}{6}
        \Cut{90}{90}{90}{10}
      \end{picture}
&
      \begin{picture}(80,80)(0,0)
        \Vertex(80,80){1.4}
        \Vertex(80,20){1.4}
        \Vertex(30,80){1.4}
        \Vertex(30,20){1.4}
        \Vertex(60,80){1.4}
        \Vertex(60,20){1.4}
        \Line(20,80)(30,80)
        \ArrowLine(30,80)(60,80)
        \ArrowLine(60,80)(80,80)
        \ArrowLine(80,80)(80,20)
        \Gluon(80,20)(60,20){3}{2}
        \Gluon(60,20)(30,20){3}{3}
        \Gluon(30,20)(20,20){3}{1}
        \Gluon(80,80)(140,80){3}{6}
        \ArrowLine(80,20)(140,20)
        \Gluon(30,80)(60,20){3}{6}
        \Gluon(60,80)(30,20){3}{6}
        \Cut{70}{90}{70}{10}
      \end{picture}
\\
      \begin{picture}(80,80)(0,0)
        \Vertex(30,80){1.4}
        \Vertex(30,20){1.4}
        \Vertex(80,80){1.4}
        \Vertex(80,20){1.4}
        \Vertex(60,20){1.4}
        \Line(20,80)(30,80)
        \ArrowLine(30,80)(80,80)
        \ArrowLine(80,80)(80,20)
        \ArrowLine(80,20)(60,20)
        \ArrowLine(60,20)(30,20)
        \Line(30,20)(20,20)
        \Gluon(80,20)(140,20){3}{6}
        \Gluon(80,80)(110,80){3}{3}
        \Vertex(110,80){1.4}
        \Gluon(110,80)(140,80){3}{3}
        \Gluon(30,80)(30,20){3}{6}
        \Gluon(60,20)(110,80){3}{7}
        \Cut{50}{90}{50}{10}
      \end{picture}
&
      \begin{picture}(100,80)(0,0)
        \Vertex(30,80){1.4}
        \Vertex(30,20){1.4}
        \Vertex(80,80){1.4}
        \Vertex(80,20){1.4}
        \Vertex(60,80){1.4}
        \Line(20,80)(30,80)
        \ArrowLine(30,80)(60,80)
        \ArrowLine(60,80)(80,80)
        \ArrowLine(80,80)(80,20)
        \ArrowLine(80,20)(30,20)
        \Line(30,20)(20,20)
        \Gluon(80,80)(140,80){3}{6}
        \Gluon(80,20)(110,20){3}{3}
        \Vertex(110,20){1.4}
        \Gluon(110,20)(140,20){3}{3}
        \Gluon(30,80)(30,20){3}{6}
        \Gluon(60,80)(110,20){3}{7}
        \Cut{50}{90}{50}{10}
      \end{picture}
&
      \begin{picture}(80,80)(0,0)
        \Vertex(30,80){1.4}
        \Vertex(30,20){1.4}
        \Vertex(80,80){1.4}
        \Vertex(80,20){1.4}
        \Vertex(60,20){1.4}
        \Line(20,80)(30,80)
        \ArrowLine(30,80)(80,80)
        \ArrowLine(80,80)(80,20)
        \Gluon(80,20)(60,20){3}{2}
        \Gluon(60,20)(30,20){3}{3}
        \Gluon(30,20)(20,20){3}{1}
        \ArrowLine(80,20)(140,20)
        \Gluon(80,80)(110,80){3}{3}
        \Vertex(110,80){1.4}
        \Gluon(110,80)(140,80){3}{3}
        \Gluon(30,80)(30,20){3}{6}
        \Gluon(60,20)(110,80){3}{7}
        \Cut{50}{90}{50}{10}
      \end{picture}
&
      \begin{picture}(80,80)(0,0)
        \Vertex(30,80){1.4}
        \Vertex(30,20){1.4}
        \Vertex(80,80){1.4}
        \Vertex(80,20){1.4}
        \Vertex(60,80){1.4}
        \Line(20,80)(30,80)
        \ArrowLine(30,80)(60,80)
        \ArrowLine(60,80)(80,80)
        \ArrowLine(80,80)(80,20)
        \Gluon(80,20)(30,20){3}{5}
        \Gluon(30,20)(20,20){3}{1}
        \Gluon(80,80)(140,80){3}{6}
        \ArrowLine(80,20)(110,20)
        \Vertex(110,20){1.4}
        \ArrowLine(110,20)(140,20)
        \Gluon(30,80)(30,20){3}{6}
        \Gluon(60,80)(110,20){3}{7}
        \Cut{50}{90}{50}{10}
      \end{picture}
\\
      \begin{picture}(80,80)(0,0)
        \Vertex(50,80){1.4}
        \Vertex(80,80){1.4}
        \Vertex(80,20){1.4}
        \Vertex(120,80){1.4}
        \Vertex(120,20){1.4}
        \Vertex(100,20){1.4}
        \ArrowLine(20,80)(50,80)
        \ArrowLine(50,80)(80,80)
        \ArrowLine(80,80)(80,20)
        \ArrowLine(80,20)(20,20)
        \Gluon(80,80)(120,80){3}{4}
        \Gluon(120,80)(140,80){3}{2}
        \Gluon(80,20)(100,20){3}{2}
        \Gluon(100,20)(120,20){3}{2}
        \Gluon(120,20)(140,20){3}{2}
        \Gluon(50,80)(100,20){3}{8}
        \Gluon(120,80)(120,20){3}{6}
        \Cut{110}{90}{110}{10}
      \end{picture}
&
      \begin{picture}(100,80)(0,0)
        \Vertex(80,80){1.4}
        \Vertex(80,20){1.4}
        \Vertex(50,20){1.4}
        \Vertex(120,80){1.4}
        \Vertex(120,20){1.4}
        \Vertex(100,80){1.4}
        \ArrowLine(20,80)(80,80)
        \ArrowLine(80,80)(80,20)
        \ArrowLine(80,20)(50,20)
        \ArrowLine(50,20)(20,20)
        \Gluon(80,20)(120,20){3}{4}
        \Gluon(120,20)(140,20){3}{2}
        \Gluon(80,80)(100,80){3}{2}
        \Gluon(100,80)(120,80){3}{2}
        \Gluon(120,80)(140,80){3}{2}
        \Gluon(120,80)(120,20){3}{6}
        \Gluon(50,20)(100,80){3}{8}
        \Cut{110}{90}{110}{10}
      \end{picture}
&
      \begin{picture}(80,80)(0,0)
        \Vertex(50,80){1.4}
        \Vertex(80,80){1.4}
        \Vertex(80,20){1.4}
        \Vertex(120,80){1.4}
        \Vertex(120,20){1.4}
        \Vertex(100,20){1.4}
        \ArrowLine(20,80)(50,80)
        \ArrowLine(50,80)(80,80)
        \ArrowLine(80,80)(80,20)
        \Gluon(80,20)(20,20){3}{6}
        \Gluon(80,80)(120,80){3}{4}
        \Gluon(120,80)(140,80){3}{2}
        \ArrowLine(80,20)(100,20)
        \ArrowLine(100,20)(120,20)
        \ArrowLine(120,20)(140,20)
        \Gluon(50,80)(100,20){3}{8}
        \Gluon(120,80)(120,20){3}{6}
        \Cut{110}{90}{110}{10}
      \end{picture}
&
      \begin{picture}(80,80)(0,0)
        \Vertex(80,80){1.4}
        \Vertex(80,20){1.4}
        \Vertex(50,20){1.4}
        \Vertex(120,80){1.4}
        \Vertex(120,20){1.4}
        \Vertex(100,80){1.4}
        \ArrowLine(20,80)(80,80)
        \ArrowLine(80,80)(80,20)
        \Gluon(80,20)(50,20){3}{3}
        \Gluon(50,20)(20,20){3}{3}
        \ArrowLine(80,20)(120,20)
        \ArrowLine(120,20)(140,20)
        \Gluon(80,80)(100,80){3}{2}
        \Gluon(100,80)(120,80){3}{2}
        \Gluon(120,80)(140,80){3}{2}
        \Gluon(120,80)(120,20){3}{6}
        \Gluon(50,20)(100,80){3}{8}
        \Cut{110}{90}{110}{10}
      \end{picture}
\\
    \end{array}
    \]
  \end{center}
  \SetScale{1.0}
  \caption{The complete set of leading $\ln s$ two-loop, two particle
    cut diagrams}
  \label{fig6}
\end{figure}
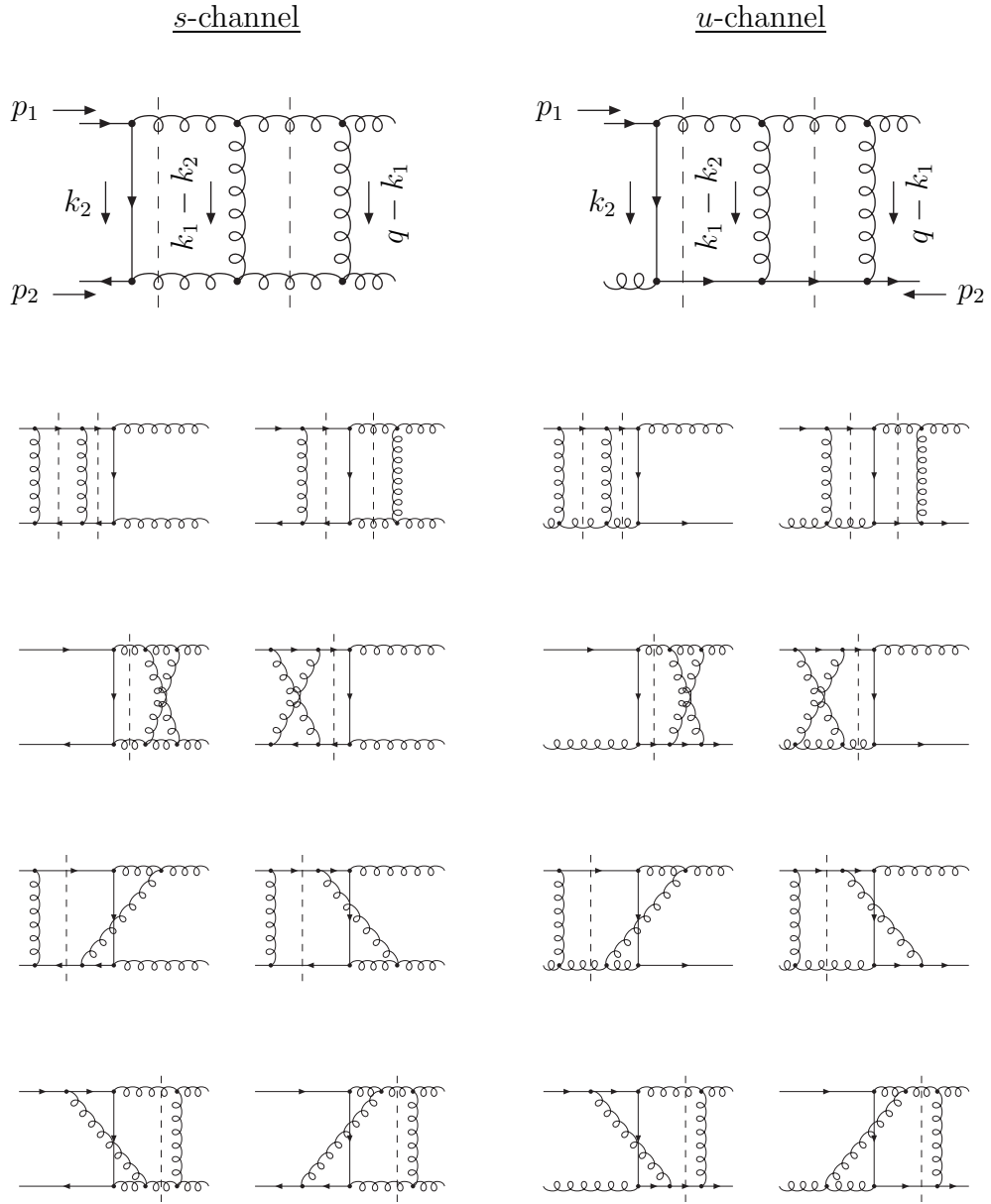

Piecing all this together with the appropriate colour factors we find that
the contribution from all such diagrams may be written as
\begin{equation}
  {\cal A}_2(q)=\frac{1}{2}  \left( \epsilon_\mathcal{Q}(q) \right)^2
  {\cal A}_0(q)  + {\cal A}_2(q)^\prime
\end{equation}
where
\begin{multline}
{\cal A}_2(q)^\prime=  
-\left(C_F+\frac{N}{2}\right)  C_F(\tau^b\tau^a)_{ij} 
 \frac{\alpha_s^3}{4\pi^3}\lns{2}\vbar(p_2)
    \gamma u(p_1)\epsilon_2^*\epsilon_1^*  \nonumber \\
 \ \times  \   \int
  d^2\vec{k_1}d^2\vec{k_2}\frac{1}{|k_2|^2|q-k_1|^2|k_1-k_2|^2} 
  + \hc  \label{eq:two-loop-main3}
\end{multline}
The first term suggests that the amplitude is indeed ``reggeizing'',
whereas the contribution ${\cal A}_2(q)^\prime$ cancels against the
diagrams shown in Fig.~\ref{fig6}, in which there are two particles in
each ($s$-channel or $u$-channel) cut and where the loop on either
side of this cut is interpreted as the reggeization of the vertical
quark or gluon line.
We have now developed a general mechanism for the derivation of the
reggeized quark, using the traditional ``bootstrap'' approach.  We
make the ansatz that the quark reggeizes, with Regge trajectory given
by $1/2+\epsilon_\mathcal{Q}(q)$ as defined in
(\ref{eq:epsilon}) and show that this ansatz is self-consistent (the
summand of $1/2$ will be explained shortly). We also need to
make use of the reggeization of the gluon which has a Regge trajectory
of $1+\epsilon_\mathcal{G}(q)$ with
\begin{equation}
  \epsilon_\mathcal{G}(q)
  =-\frac{N}{2}\frac{\alpha_s}{2\pi^2} \int d^2\vec{k}
 \frac{|q|^2}{|k|^2 |k-q|^2 }.
\end{equation}

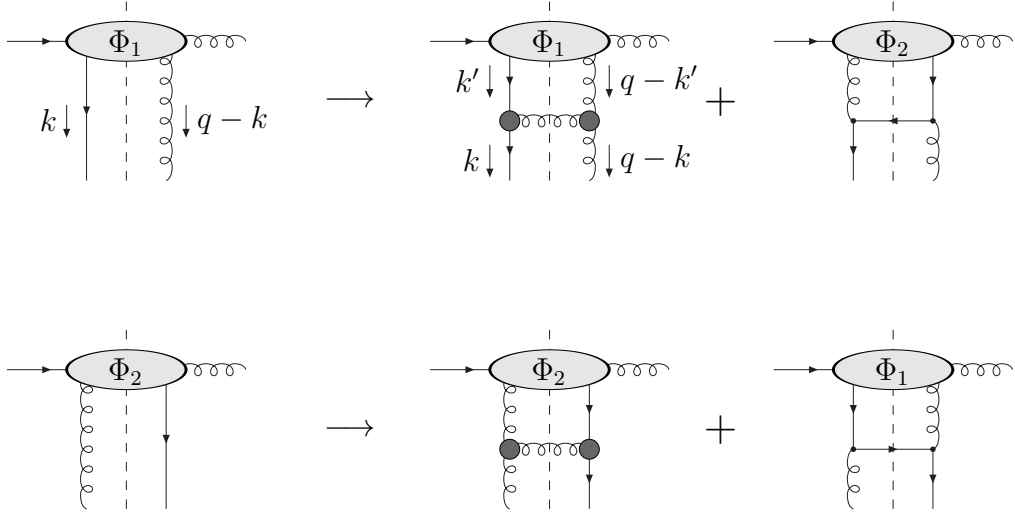
\begin{figure}
  \begin{center}
    \leavevmode
    \SetScale{0.75}
    \[
    \begin{array}{ccc}
      \begin{picture}(120,90)(20,0)
        \ArrowLine(20,90)(60,90)
        \ArrowLine(60,90)(60,20)
        \Gluon(100,90)(140,90){3}{4}
        \Gluon(100,90)(100,20){3}{6}
        \Cut{80}{110}{80}{20}
        \GOval(80,90)(10,30)(0){.9}
        \Text(60,67)[c]{${\Phi_1}$}
        \LongArrow(50,58)(50,43)
        \Text(34,38)[r]{$k$}
        \LongArrow(110,58)(110,43)
        \Text(87,38)[l]{$q-k$}
        \Text(135,45)[l]{$\mbox{{$\longrightarrow$}}$}
      \end{picture}
&
      \begin{picture}(120,90)(-10,0)
        \ArrowLine(20,90)(60,90)
        \ArrowLine(60,90)(60,50)
        \ArrowLine(60,50)(60,20)
        \Gluon(100,90)(140,90){3}{4}
        \Gluon(100,90)(100,20){3}{6}
        \Gluon(60,50)(100,50){3}{4}
        \Cut{80}{110}{80}{20}
        \GOval(80,90)(10,30)(0){.9}
        \Text(60,67)[c]{${\Phi_1}$}
        \LongArrow(50,38)(50,23)
        \Text(34,23)[r]{$k$}
        \LongArrow(50,78)(50,63)
        \Text(34,53)[r]{$k'$}
        \LongArrow(110,38)(110,23)
        \Text(87,23)[l]{$q-k$}
        \LongArrow(110,78)(110,63)
        \Text(87,53)[l]{$q-k'$}
        \GOval(60,50)(5,5)(0){.4} \GOval(100,50)(5,5)(0){0.4}
        \Text(125,45)[c]{\Large{+}}
      \end{picture}
&
      \begin{picture}(120,90)(-10,0)
        \ArrowLine(20,90)(60,90)
        \ArrowLine(100,90)(100,50)
        \ArrowLine(60,50)(60,20)
        \Gluon(100,90)(140,90){3}{4}
        \Gluon(60,50)(60,90){3}{3}
        \Gluon(100,50)(100,20){3}{2}
        \ArrowLine(100,50)(60,50)
        \Cut{80}{110}{80}{20}
        \GOval(80,90)(10,30)(0){.9}
        \Vertex(100,50){1.4}
        \Vertex(60,50){1.4}
        \Text(60,67)[c]{${\Phi_2}$}
      \end{picture}
\\
      \begin{picture}(120,120)(20,0)
        \ArrowLine(20,90)(60,90)
        \ArrowLine(100,90)(100,20)
        \Gluon(100,90)(140,90){3}{4}
        \Gluon(60,20)(60,90){3}{6}
        \Cut{80}{110}{80}{20}
        \GOval(80,90)(10,30)(0){.9}
        \Text(60,67)[c]{${\Phi_2}$}
        \Text(135,45)[l]{$\mbox{{$\longrightarrow$}}$}
      \end{picture}
&
      \begin{picture}(120,120)(-10,0)
        \ArrowLine(20,90)(60,90)
        \ArrowLine(100,90)(100,50)
        \ArrowLine(100,50)(100,20)
        \Gluon(100,90)(140,90){3}{4}
        \Gluon(60,50)(60,90){3}{3}
        \Gluon(60,20)(60,50){3}{2}
        \Gluon(60,50)(100,50){3}{4}
        \Cut{80}{110}{80}{20}
        \GOval(80,90)(10,30)(0){.9}
        \Text(60,67)[c]{${\Phi_2}$}
        \GOval(60,50)(5,5)(0){.4} \GOval(100,50)(5,5)(0){0.4} 
        \Text(125,45)[c]{\Large{+}}
      \end{picture}
&
      \begin{picture}(120,120)(-10,0)
        \ArrowLine(20,90)(60,90)
        \ArrowLine(60,90)(60,50)
        \ArrowLine(100,50)(100,20)
        \Gluon(100,90)(140,90){3}{4}
        \Gluon(100,90)(100,50){3}{3}
        \Gluon(60,20)(60,50){3}{2}
        \ArrowLine(60,50)(100,50)
        \Cut{80}{110}{80}{20}
        \GOval(80,90)(10,30)(0){.9}
        \Vertex(100,50){1.4}
        \Vertex(60,50){1.4}
        \Text(60,67)[c]{${\Phi_1}$}
      \end{picture}      
    \end{array}
    \]
  \end{center}
    \caption{The addition of a rung involves the consideration of 
      both a gluon rung with effective vertices and of a quark rung.
      Only the upper-half of the diagrams is shown, and note that the
      quark rung diagrams imply that we must have $\Phi_1=\Phi_2$}
  \label{fig7}
\end{figure}

In order to develop an expression for the amplitude for the process
(\ref{process1}), whose iterative solution gives a series
$$
{\cal A}_0(q) + {\cal A}_1(q) + {\cal A}_2(q) + \cdots,
$$
we need to consider the addition of both gluon \emph{and} quark or
antiquark rungs. This is demonstrated in Fig.~\ref{fig7}.
In general these ladder diagrams lead to an integro-differential
equation of the form
\begin{equation}
  \frac{\partial F(s,k,q)}{\partial \ln s}=
  \int d^2\vec{k'}\,\mathcal{K}(k,k',q)F(s,k',q),
\end{equation}
in which the kernel, $\mathcal{K}$, may be divided into two parts. The
effect of adding a rung gives a contribution of
\begin{equation}
 \mathcal{K}^{{\rm rung}}(k,k',q)=-\frac{\alpha_s}{2\pi^2}
 \frac{k'}{|k'|^2|q-k'|^2}
 \left(C_F q^* - C_F\frac{{k}^*|q-k'|^2}{|k-k'|^2}
   -\frac{N}{2}\frac{{k'}^*|q-k|^2}{|k-k'|^2}\right) + \hc
\end{equation}
The first term is precisely what one would expect from the
reggeization of the quark, whereas the second term cancels when we
take into account the fact that the quark and gluon exchanged in the
$t$-channel must themselves be reggeized and this reggeization gives
rise to a correction
\begin{equation}
  \int d^2\vec{k'}\,\mathcal{K}^{{\rm
      regge}}(k,k',q)=\left(\epsilon_\mathcal{Q}(k)+
    \epsilon_\mathcal{G}(q-k)\right).
\end{equation}

When all is pieced together we find that the effects of adding a rung
\emph{and} of taking into account the reggeization of the quarks and
gluons exchanged in the $t$-channel is such that 
\begin{equation}
  \int d^2\vec{k'}\,(\mathcal{K}^{\rm rung}(k,k',q)+\mathcal{K}^{\rm
    regge}(k,k',q)) =\epsilon_\mathcal{Q}(q),
\end{equation}
which means that the leading $\ln s$ part of the amplitude, summed to
all orders in perturbation theory, gives rise to an amplitude of the
form
\begin{equation}
  s^{\epsilon_\mathcal{Q}(q)} {\cal A}_0 \ \sim \ 
  s^{1/2+\epsilon_\mathcal{Q}(q)},
\end{equation}
where the ``na\"{\i}ve'' $s$ dependence $s^{1/2}$ arises from the usual
$s$ dependence of an amplitude in which a spin-$\frac{1}{2}$
particle is exchanged in the $t$-channel and is contained in the normalisation
of the spinors in (\ref{eq:tree-level2}).


\section{The Integral Equation for the Reggeon Amplitude and its Solution}

The integral equation for the perturbative amplitude for processes
involving the exchange of a Reggeon are obtained by considering
ladders in which the vertical lines are a reggeized quark-antiquark
pair, and the horizontal rungs are gluons which couple to these
reggeized quarks via the effective vertex given in
(\ref{quarkeffective}). The colour singlet is projected from the two
fermions exchanged in the $t$-channel.  This has been studied in
\cite{kirschner}. One obtains an integral equation for the Mellin
transform of the amplitude, $f(s,k,q)$ defined by
$$
\tilde{f}(\omega,k,q) = \int \frac{ds}{s} \left( \frac{s}{\vec{k}^2} \right)^{-\omega} f(s,k,q), $$
where $k$ is the transverse
momentum of the incoming quark and $q$ is the momentum transfer (which
may be taken to be transverse).  In terms of this Mellin transform one
obtains
\begin{multline}
  \label{eq:kirschner1}
  \tilde f(\omega,k,q) = \tilde f_0(\omega,k,q)
  +\frac{\tilde\alpha_s}{2\pi\omega}\int
  d^2\vec{k'}\left\{
    \frac{\vec{k'}\cdot(\vec{k'}-\vec{q})}{\vec{k'}^2(\vec{k'}-\vec{q})^2}
    \left[\min\left(\frac{k}{k'},\frac{k'}{k}\right)\right]^\omega\right. 
  \\
  +\left.\frac{\vec{k'}\cdot\vec{k}}{\vec{k'}^2(\vec{k}-\vec{k'})^2}
  +\frac{(\vec{k}-\vec{q})\cdot(\vec{k'}-\vec{q})}
  {(\vec{k'}-\vec{q})^2(\vec{k}-\vec{k'})^2}\right\} \tilde
  f(\omega,k',q) \\
  -\frac{\tilde\alpha_s}{2\pi\omega}\int
  d^2\vec{k'}\left\{\frac{\vec{k'}\cdot\vec{k}}{\vec{k'}^2(\vec{k}-\vec{k'})^2}
  +\frac{(\vec{k}-\vec{q})\cdot(\vec{k'}-\vec{q})}
  {(\vec{k'}-\vec{q})^2(\vec{k}-\vec{k'})^2}\right\} \tilde
  f(\omega,k,q),
\end{multline}
where we have used the notation
$$
\tilde\alpha_s \ = \ C_F \frac{\alpha_s}{\pi},
$$
and $\tilde f_0(\omega,k,q)$ is the Born term obtained from the leading order
contribution consisting of the exchange of a quark-antiquark pair
(with no gluons).

There is an important difference between this equation and the
equivalent equation \cite{bfkl} for the Pomeron, which is built out of
reggeized gluons in a similar fashion. In the case of the Pomeron a
solution which is of the reggeized form (i.e. a simple pole in the
Mellin transform at a value of $\omega$ equal to the Regge trajectory)
is prevented by the fact that the colour factor which arises in the
projection of a colour singlet from the two gluons in the $t$-channel
is twice that obtained from the projection of a colour octet (relevant
for the integral equation describing the reggeized gluon itself).  In
the case of the Reggeon, a colour factor of $C_F$ is obtained
\emph{both} in the case of the reggeized gluon, and in the projection
of a colour singlet from a quark-antiquark pair.  In the Reggeon case,
simple reggeization is prevented by the $\omega$ dependent factor
$$
\left[\min\left(\frac{k}{k'},\frac{k'}{k}\right)\right]^\omega,
$$
which occurs inside the kernel on the RHS of 
(\ref{eq:kirschner1}). This term arises from the fact that when one
exchanges fermions in the $t$-channel such that the momenta of these fermions
appear in the numerators of the propagators, care must be taken 
in setting the kinematic limits of the phase-space integral over the
 longitudinal components of momentum \cite{kirschner}. It is these
kinematic limits which lead to the  $\omega$ dependent term in the kernel
and, as we shall see later, this has a radical effect on the solution
to this equation.

Henceforth we shall be confining our discussion to the case of zero momentum
transfer ($q=0$), for which for which  the second term
on the RHS of (\ref{eq:kirschner1}) simplifies
to
\begin{multline}
  \label{eq:aaaap}
  \tilde \mathcal{K}(\omega,k,k')\otimes \tilde f(\omega,k')=
  \frac{\tilde{\alpha}_s}{2}
  \int\frac{d{k'}^2}{{k'}^2}\left\{\left[ \min\left(\frac{k}{k'},
  \frac{k'}{k}\right)^\omega -1\right]\tilde f(\omega,k') \right. 
  \\
  \left.+ \frac{k^2+{k'}^2}{|k^2-{k'}^2|}\left(\tilde f(\omega,k')-\tilde f(\omega,k)
  \right) + \tilde f(\omega,k)\right\},
\end{multline}
where we have integrated over the angular part of $k$, assuming that
the Reggeon is dominated by an amplitude which is azimuthally symmetric,
i.e. that $ \tilde f(\omega,k')$ depends only on the magnitude,
 $|k^\prime|$.

This can be solved exactly by exploiting the two-dimensional
 conformal invariance of the kernel to conclude that a function
of the form
$$ 
\tilde f(\omega, k ) = \left( |k|^2 \right)^{i\nu}
$$
is an eigenfunction of the kernel with eigenvalue $\omega(\nu)$
given by the solution to the equation
\begin{equation}
  \omega=
  \frac{\tilde{\alpha}_s}{2}\left[\frac{\omega}{\omega^2/4+\nu^2}-2\psi(1+i\nu)-2\psi(1-i\nu)
    -4\gamma_E\right]. \label{omega}
\end{equation}
The leading eigenvalue is given by
\footnote{ We note here that $-\omega_0$ is also an eigenvalue of the kernel,
 and although it is sub-leading, it will turn out to be convenient to exploit
 this in our subsequent analysis.}
$$
\omega_0 = \sqrt{2 \tilde{\alpha}_s},
$$
which implies that the Reggeon has an asymptotic $s$ behaviour of the form
$$
s^{\sqrt{2\tilde{\alpha}_s}}.
$$
Another qualitative difference between the QCD Reggeon and the QCD
Pomeron lies in the fact that such a leading $s$ behaviour, which is
not analytic in the coupling, cannot be reproduced in an ordinary
perturbative expansion in $\alpha_s$, although the even orders in the
expansion of this $s$ dependence matches the double leading logarithms
found in ordinary perturbation theory \cite{lipatov}.

Thus, in principle, we have an analytic method for solving the
integral equation for the Mellin transform of the Reggeon exchange
amplitude.  However, in order to be able to introduce further
refinements, such as the running of the coupling or some simulation of
non-perturbative effects which necessarily break the conformal
invariance, one needs to be able to carry out a programme of numerical
solution of (\ref{eq:kirschner1}).

This is most easily achieved by inverting the Mellin transform
to obtain an integro-differential equation
\begin{equation}
  \label{eq:kirschner_differential}
  \frac{\partial f(s ,k)}{\partial\ln s }=
  \mathcal{K}(s ,k,k')\otimes f(s ,k'),
\end{equation}
where
\begin{multline}
  \mathcal{K}(s ,k,k')\otimes f(s ,k') = \tilde{\alpha}_s
  \left\{\int_0^1 \frac{dz}{z}
  \frac{2z^2}{1-z^2}\left(
    f(s ,kz)+f(s ,k/z)-2f(s ,k)\right)\right. \\ 
  +\left.\int_{1/s }^1 \frac{dz}{z}\left(
      f(s z,kz)+f(s z,k/z)\right)\right\}
  \label{boo}
\end{multline}
and we have introduced the variable $z$ to represent
 $\min(k/k',k'/k)$. The $\omega$ dependent part of
the kernel in Mellin transform space is now encoded by the scaling of
$s$ by a factor of $z$ in the appropriate terms.

We know that the asymptotic behaviour of the solution to this
differential equation should be $s^{\omega_0}$. The usual method of
numerical solution of such differential equations is to start with an
initial function which has no $s$ dependence and to integrate in steps
using the Runge-Kutta method. Unfortunately, since the leading $s$
behaviour arises from the $\omega$ dependent part of the kernel in
Mellin transform space, which has now been transferred to the $s$
dependent part of $f(s,k)$ on the RHS of (\ref{boo}), an initial
function which does not contain any $s$ dependence can {\it never}
generate the expected asymptotic behaviour in any step-by-step
numerical integration routine.

This difficulty can be circumvented by extracting the leading
$s^{\omega_0}$ behaviour. This is achieved by defining a function
$g(s,k)$ by
\begin{equation}
  f(s,k) = \cosh \left( \omega_0 \ln(s) \right) g(s,k).
  \label{gdefined}
\end{equation}
The asymptotic behaviour of $f(s,k)$ is now displayed explicitly
and we expect the function $g(s,k)$ to have a far less dramatic
$s$ dependence. The integro-differential equation for $g(s,k)$
is given by
\begin{multline}
  \frac{\partial g(s ,k)}{\partial \ln s } =
  \mathrm{sech}(\omega_0\ln s )\mathcal{K}(s ,k,k')\otimes
  \left(\cosh(\omega_0\ln s )g(s ,k')\right) 
  -\omega_0\tanh(\omega_0\ln s )g(s ,k)
  \label{eq:kirschner_for_g}
\end{multline} 
where the kernel $\mathcal{K}(s ,k,k')$ is defined in (\ref{boo}).
We note here that $g(s,k)=\mathrm{constant}$ is an exact solution of
(\ref{eq:kirschner_for_g}). We may therefore start with a function,
$g_0(k)$, that depends only on $k$ but {\it not} on $s$ (which one
would normally obtain from an ``impact factor'' which encodes the
coupling of the Reggeon to the particles between which the Reggeon is
exchanged) and generate (after sufficient iterations) a solution for
the Reggeon exchange amplitude with initial value $g_0(k)$ taken as
the value of the amplitude at low $s$.

We have checked numerically that this method reproduces the known analytic
results in the case of a trial initial function, which has a simple Fourier
transform. We have taken
\begin{equation}
  g_0(k)  = \frac{|k|^2}{(|k|^2+1)^2} = 
  \frac{1}{\pi} \int_0^\infty d \nu \, \Gamma(1+i \nu) \Gamma(1-i \nu)
  \left( |k|^2 \right)^{i \nu}.
\end{equation} 
For large $s$ the amplitude is then given by
\begin{equation}
  f(s,k) = \frac{1}{\pi} \int_0^\infty d \nu
  \, \Gamma(1+i \nu) \Gamma(1-i \nu) \left( |k|^2 \right)^{i \nu}
  s^{\omega(\nu)},
\end{equation}
where $\omega(\nu)$ is given by (\ref{omega}).  This can be compared
with the numerical method described above.  Having established the
validity of this numerical approach, one can introduce refinements to
the kernel, which render it intractable to analytical methods, but
still open to numerical analysis. In the next section we discuss the
results of such an approach.


\section{Numerical Results}

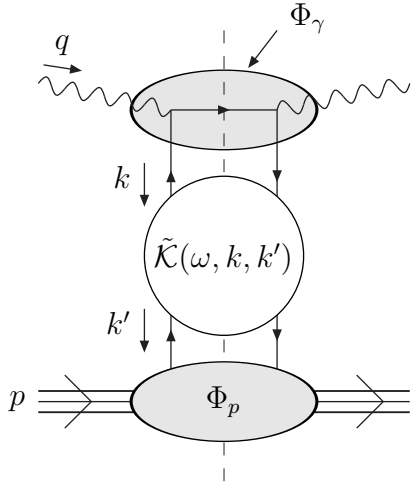
\begin{figure}
  \begin{center}
    \begin{picture}(160,190)(0,-60)
      \GOval(80,90)(15,35)(0){.9}
      \Text(105,125)[l]{${\Phi}_\gamma$}
      \LongArrow(100,120)(90,107)
      \Photon(10,100)(60,90){3}{5}
      \ArrowLine(60,-20)(60,30)
      \ArrowLine(100,90)(100,40)
      \ArrowLine(60,40)(60,90)
      \ArrowLine(100,30)(100,-20)
      \ArrowLine(60,90)(100,90)
      \Photon(100,90)(150,100){3}{5}

      \SetWidth{0.7}
      \Line(10,-24)(150,-24)
      \Line(10,-16)(150,-16)
      \SetWidth{0.5}
      \Line(10,-20)(150,-20)
      \Line(20,-10)(30,-20)
      \Line(30,-20)(20,-30)
      \Line(130,-10)(140,-20)
      \Line(140,-20)(130,-30)

      \Cut{80}{120}{80}{-50}
      
      \LongArrow(50,70)(50,55)
      \Text(45,65)[r]{$k$}
      \LongArrow(50,15)(50,0)
      \Text(45,10)[r]{$k'$}
      \LongArrow(12,107)(25,105)
      \Text(19,115)[c]{$q$}
      \Text(5,-20)[r]{$p$}

      \GOval(80,35)(30,30)(0){1}
      \GOval(80,-20)(15,35)(0){.9}
      \Text(80,-20)[c]{$\Phi_p$}
      \Text(80,35)[c]{$\tilde\mathcal{K}(\omega,k,k')$}
    \end{picture}
  \end{center}
  \caption{The impact factors $\Phi_\gamma$ and $\Phi_p$ encode how
    the virtual photon (or $W$-, $Z$-boson) and the proton
    respectively couple to the Reggeized quark ladder. The central
    circle symbolises the kernel's role in adding rungs and Reggeizing
    the uprights}
  \label{fig8}
\end{figure}

In this section we apply the technique discussed in the last section
to a model of deep-inelastic scattering processes at low-$x$, which
would be dominated by Reggeon rather than Pomeron exchange (such as
the structure function $F_3$ for virtual $W$- or $Z$-boson scattering,
or the spin-dependent structure functions which are intrinsically
flavour non-singlet). For the rest of this section we shall use the
term ``structure function'' to mean the flavour non-singlet part of a
structure function, which we expect to be dominated by Reggeon
exchange at low-$x$.
\begin{figure}
  \begin{center}
\setlength{\unitlength}{0.1bp}
\begin{picture}(3600,2160)(0,0)
\put(2150,150){\makebox(0,0){$x$}}
\put(100,1230){%
\makebox(0,0)[b]{\shortstack{$f(x,Q^2)$}}%
}
\put(750,300){\makebox(0,0){$10^{-4}$}}
\put(2617,300){\makebox(0,0){$10^{-2}$}}
\put(1683,300){\makebox(0,0){$10^{-3}$}}
\put(3550,300){\makebox(0,0){$10^{-1}$}}
\put(700,660){\makebox(0,0)[r]{$10^{-2}$}}
\put(700,1995){\makebox(0,0)[r]{$10^0$}}
\put(700,1328){\makebox(0,0)[r]{$10^{-1}$}}
\end{picture}
    \caption
    {The (lack of) dependence of results on the parameter $\eta$ in
      (\ref{phi2}) is demonstrated. The structure function at fixed
      $Q^2$ is given for six values of $\eta$ from 0.6 to 2.0}
    \label{fig9}
  \end{center}
\end{figure}
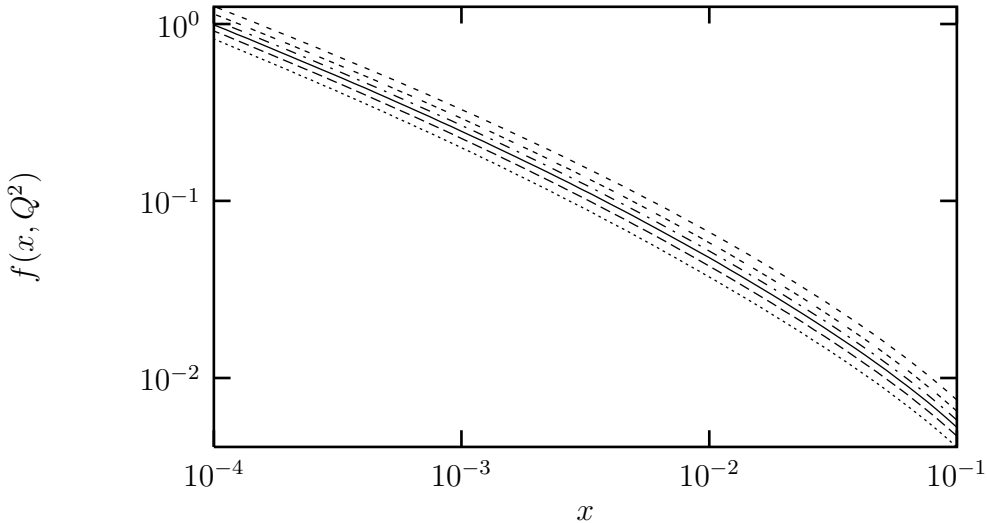
In order to do this we need some model for the ``impact factors'' which account
for the coupling of the Reggeon at the top and bottom of the ladder,
as indicated in Fig.~\ref{fig8}. Unlike the case of the Pomeron,
there is a direct coupling of the quarks in the Reggeon to the virtual
particle involved in the scattering (photon, $W$-boson, or $Z$-boson).
This means that the upper impact factor is proportional to
$$ \delta(k^2-Q^2),$$
where $-Q^2$ is the square momentum of the virtual boson. At the other end
 of the ladder we need the impact factor for the coupling of the Reggeon to the target proton. This is non-perturbative in nature and so we model this with
the simple form
\begin{equation}
  \Phi_p \propto e^{-k^2/\eta \Lambda_{\mathrm{QCD}}^2},
  \label{phi2}
\end{equation}
where we expect $\eta$ to be of order unity, and we have not fixed the
constant of proportionality which means that our scale remains
arbitrary. This is clearly a crude estimate of the lower impact
factor, but we do not expect our results for the low-$x$, large $Q^2$
regime to depend critically on the exact form of this impact factor.
This is because after a sufficient number of iterations of the kernel
one expects the amplitude to lose all ``memory'' of the initial
condition. As a demonstration of this, we show in Fig.~\ref{fig9} the
variation of the structure function plotted against $x$ for $Q^2 \sim
100 \ {\rm GeV^2}$ for a range of the parameter $\eta$ between 0.6 and
2.0. We see that the variation of the shape of the structure is indeed
small and the differences can be substantially absorbed into an
overall normalization.

We start by assuming fixed coupling. In this case the behaviour of the
structure function at low $x$ for each value of $Q^2$ was obtained by
starting with an initial function given by (\ref{phi2}) and
numerically integrating (\ref{eq:kirschner_for_g}) using the
Runge-Kutta method with the coupling $\tilde{\alpha}_s$ taken to be
$\tilde{\alpha}_s(Q^2)$. From the resulting function $f(x,k)$ ($s$ is
replaced by $1/x$), the point $k^2=Q^2$ was selected, thereby imposing
the delta-function which represents the impact factor at the top of
the ladder. This process was then repeated for different values of
$Q^2$.

Next we wish to include the effect of the running of the coupling.
This is achieved by promoting the coupling $\tilde{\alpha}_s$ in the
expression for the kernel to a running coupling. We take the larger of
the momenta $k$ and $k^\prime$ as the argument of the running of the
coupling, which means that the coupling must now be taken
\emph{inside} the integral in (\ref{boo}). One immediate effect of
this is that the quantity $\omega_0$, which was the leading eigenvalue
of the kernel in the case of fixed coupling, now becomes a function of
$k$ given by
\begin{equation}
  \omega_0(k)=\frac{1}{2}\int\frac{d{k'}^2}{{k'}^2}
  \tilde{\alpha}_s(\max(k,{k'}))
  \left[\min\left(\frac{k}{k'},\frac{k'}{k}\right)\right]^{\omega_0(k^2)},
\end{equation}
and it is this value of $\omega(k)$ that is used to factor off
the leading behaviour. Thus we define $g(x,k)$ as
$$
f(x,k) = \cosh \left( \omega(k) \ln(s) \right) g(x,k),
$$
and the $x$ dependence of $g(x,k)$ is given by
(\ref{eq:kirschner_for_g}) with $\tilde{\alpha}_s$ replaced by
$\tilde{\alpha}_s(\max(k,k^\prime))$.  Note that in this case, unlike
the case of fixed coupling, we only need to solve the corresponding
differential equation once, since the argument of the coupling for
each iteration does \emph{not} now depend on $Q^2$. A single pass
through the range of $x$ returns a function $f(x,k)$ and the results
for the structure function at any value of $Q^2$ can be read off by
setting $k^2=Q^2$.

As is always the case when we introduce a running coupling in such
integro-differential equations, we need to impose an infra-red cutoff
below which $\alpha_s$ takes the value $\alpha_{\max}$ and ceases to run.
This is because the integral in (\ref{boo}) samples {\it all}
possible  momenta so that without the imposition of such an infra-red
cutoff one would run into the Landau pole. We have chosen $\alpha_{\max}$
to take the value 1.  
\begin{figure}[p]
  \begin{center}
\setlength{\unitlength}{0.1bp}
\begin{picture}(3600,2160)(0,0)
\put(3137,1647){\makebox(0,0)[r]{$Q=80$}}
\put(3137,1747){\makebox(0,0)[r]{$Q=30$}}
\put(3137,1847){\makebox(0,0)[r]{$Q=10$}}
\put(3137,1947){\makebox(0,0)[r]{$Q=2$}}
\put(1287,874){\makebox(0,0)[l]{$\propto x^{-1/2}$}}
\put(2075,150){\makebox(0,0){$x$}}
\put(100,1230){%
\makebox(0,0)[b]{\shortstack{$f(x,Q^2)$}}%
}
\put(3550,300){\makebox(0,0){$10^{-2}$}}
\put(2567,300){\makebox(0,0){$10^{-3}$}}
\put(1583,300){\makebox(0,0){$10^{-4}$}}
\put(600,300){\makebox(0,0){$10^{-5}$}}
\put(550,2060){\makebox(0,0)[r]{$10^{5}$}}
\put(550,1823){\makebox(0,0)[r]{$10^{4}$}}
\put(550,1586){\makebox(0,0)[r]{$10^{3}$}}
\put(550,1349){\makebox(0,0)[r]{$10^{2}$}}
\put(550,1111){\makebox(0,0)[r]{$10^{1}$}}
\put(550,874){\makebox(0,0)[r]{$10^{0}$}}
\put(550,637){\makebox(0,0)[r]{$10^{-1}$}}
\put(550,400){\makebox(0,0)[r]{$10^{-2}$}}
\end{picture}
    \caption{The rise in the structure function at low-$x$ is shown here for four different values of
      $Q$ (in [GeV]) for static $\alpha_s$. The na\"{\i}ve dependence
      of $x^{-1/2}$ is also shown for comparison}
    \label{fig10}
  \end{center}
\end{figure}
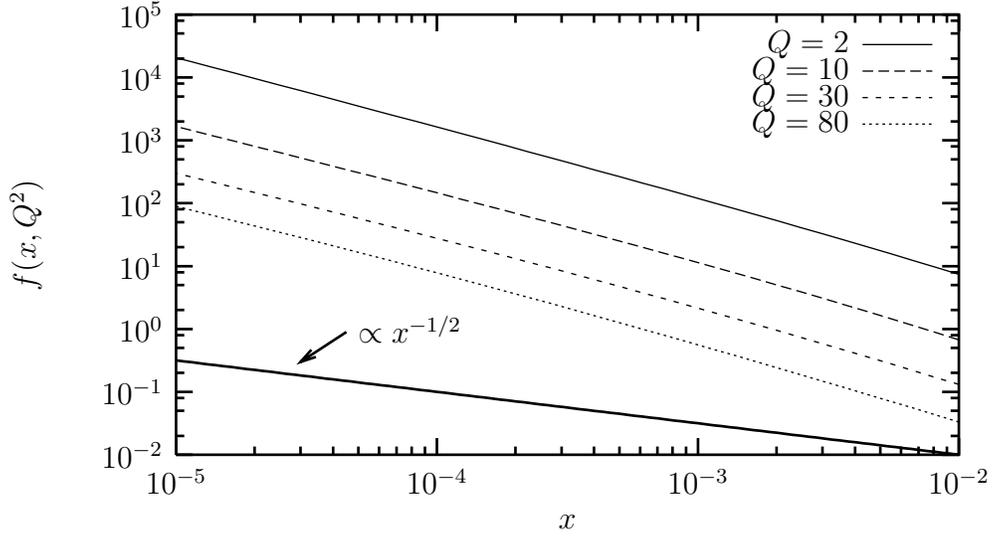
\begin{figure}[p]
  \begin{center}
\setlength{\unitlength}{0.1bp}
\begin{picture}(3600,2160)(0,0)
\put(3137,1647){\makebox(0,0)[r]{$Q=80$}}
\put(3137,1747){\makebox(0,0)[r]{$Q=30$}}
\put(3137,1847){\makebox(0,0)[r]{$Q=10$}}
\put(3137,1947){\makebox(0,0)[r]{$Q=2$}}
\put(1287,920){\makebox(0,0)[l]{$\propto x^{-1/2}$}}
\put(2075,150){\makebox(0,0){$x$}}
\put(100,1230){%
\makebox(0,0)[b]{\shortstack{$f(x,Q^2)$}}%
}
\put(3550,300){\makebox(0,0){$10^{-2}$}}
\put(2567,300){\makebox(0,0){$10^{-3}$}}
\put(1583,300){\makebox(0,0){$10^{-4}$}}
\put(600,300){\makebox(0,0){$10^{-5}$}}
\put(550,2060){\makebox(0,0)[r]{$10^{6}$}}
\put(550,1783){\makebox(0,0)[r]{$10^{5}$}}
\put(550,1507){\makebox(0,0)[r]{$10^{4}$}}
\put(550,1230){\makebox(0,0)[r]{$10^{3}$}}
\put(550,953){\makebox(0,0)[r]{$10^{2}$}}
\put(550,677){\makebox(0,0)[r]{$10^{1}$}}
\put(550,400){\makebox(0,0)[r]{$10^{0}$}}
\end{picture}
    \caption{The same as figure~\ref{fig10} but with a running
    coupling. Note that the low-$x$ rise is now even more pronounced}
    \label{fig11}
  \end{center}
\end{figure}
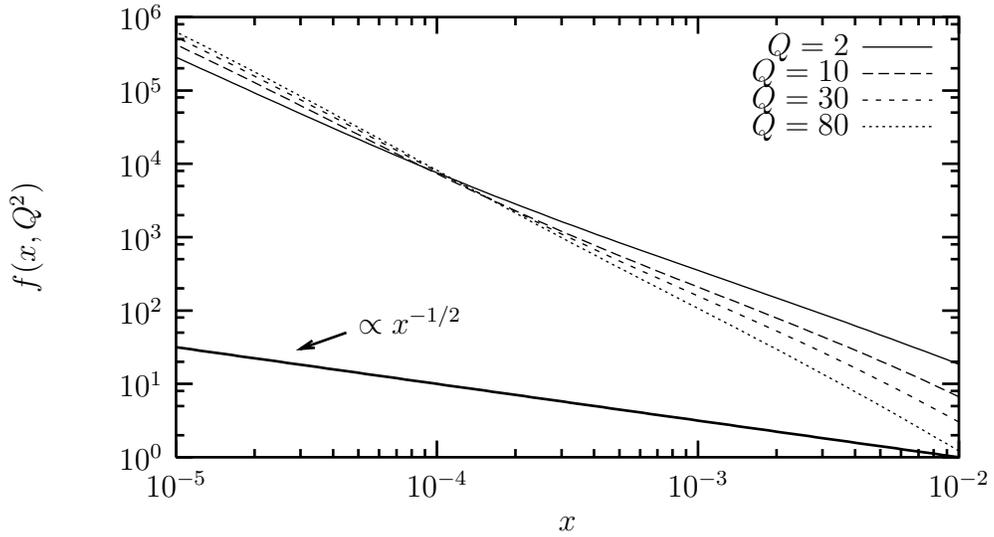
\begin{figure}
  \begin{center}
\setlength{\unitlength}{0.1bp}
\begin{picture}(3600,2160)(0,0)
\put(3137,1947){\makebox(0,0)[r]{$x=10^{-4}$}}
\put(3137,1847){\makebox(0,0)[r]{$x=10^{-3}$}}
\put(3137,1747){\makebox(0,0)[r]{$x=10^{-2}$}}
\put(2050,150){\makebox(0,0){$Q$ [GeV]}}
\put(100,1230){%
\makebox(0,0)[b]{\shortstack{$f(x,Q^2)$}}%
}
\put(3550,300){\makebox(0,0){90}}
\put(3217,300){\makebox(0,0){80}}
\put(2883,300){\makebox(0,0){70}}
\put(2550,300){\makebox(0,0){60}}
\put(2217,300){\makebox(0,0){50}}
\put(1883,300){\makebox(0,0){40}}
\put(1550,300){\makebox(0,0){30}}
\put(1217,300){\makebox(0,0){20}}
\put(883,300){\makebox(0,0){10}}
\put(550,300){\makebox(0,0){0}}
\put(500,2060){\makebox(0,0)[r]{$10^{4}$}}
\put(500,1783){\makebox(0,0)[r]{$10^{3}$}}
\put(500,1507){\makebox(0,0)[r]{$10^{2}$}}
\put(500,1230){\makebox(0,0)[r]{$10^1$}}
\put(500,953){\makebox(0,0)[r]{$10^0$}}
\put(500,677){\makebox(0,0)[r]{$10^{-1}$}}
\put(500,400){\makebox(0,0)[r]{$10^{-2}$}}
\end{picture}
    \caption{The structure function $f$ shown against $Q$ for three
    values of $x$. The coupling has been fixed for this calculation}
    \label{fig12}
  \end{center}
\end{figure}
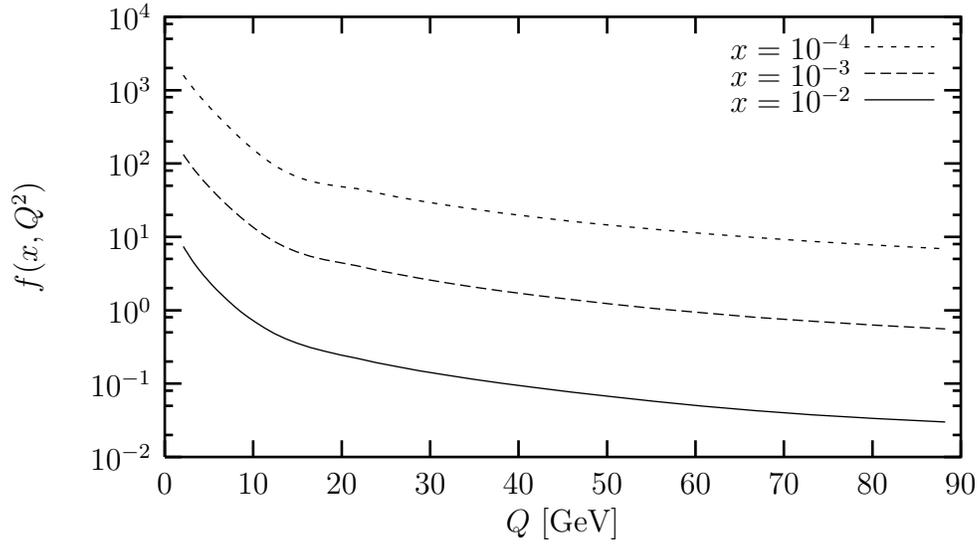
\begin{figure}
  \begin{center}
\setlength{\unitlength}{0.1bp}
\begin{picture}(3600,2160)(0,0)
\put(3137,1947){\makebox(0,0)[r]{$x=10^{-4}$}}
\put(3137,1847){\makebox(0,0)[r]{$x=10^{-3}$}}
\put(3137,1547){\makebox(0,0)[r]{$x=10^{-2}$}}
\put(2075,150){\makebox(0,0){$Q$ [GeV]}}
\put(100,1230){%
\makebox(0,0)[b]{\shortstack{$f(x,Q^2)$}}%
}
\put(3550,300){\makebox(0,0){90}}
\put(3215,300){\makebox(0,0){80}}
\put(2880,300){\makebox(0,0){70}}
\put(2544,300){\makebox(0,0){60}}
\put(2209,300){\makebox(0,0){50}}
\put(1874,300){\makebox(0,0){40}}
\put(1539,300){\makebox(0,0){30}}
\put(1203,300){\makebox(0,0){20}}
\put(868,300){\makebox(0,0){10}}
\put(550,2060){\makebox(0,0)[r]{$10^{5}$}}
\put(550,1728){\makebox(0,0)[r]{$10^{4}$}}
\put(550,1396){\makebox(0,0)[r]{$10^{3}$}}
\put(550,1064){\makebox(0,0)[r]{$10^{2}$}}
\put(550,732){\makebox(0,0)[r]{$10^{1}$}}
\put(550,400){\makebox(0,0)[r]{$10^{0}$}}
\end{picture}
    \caption{Similar to figure~\ref{fig12}, but with the
    coupling now running. This acts to reduce the rise at small $Q$}
    \label{fig13}
  \end{center}
\end{figure}
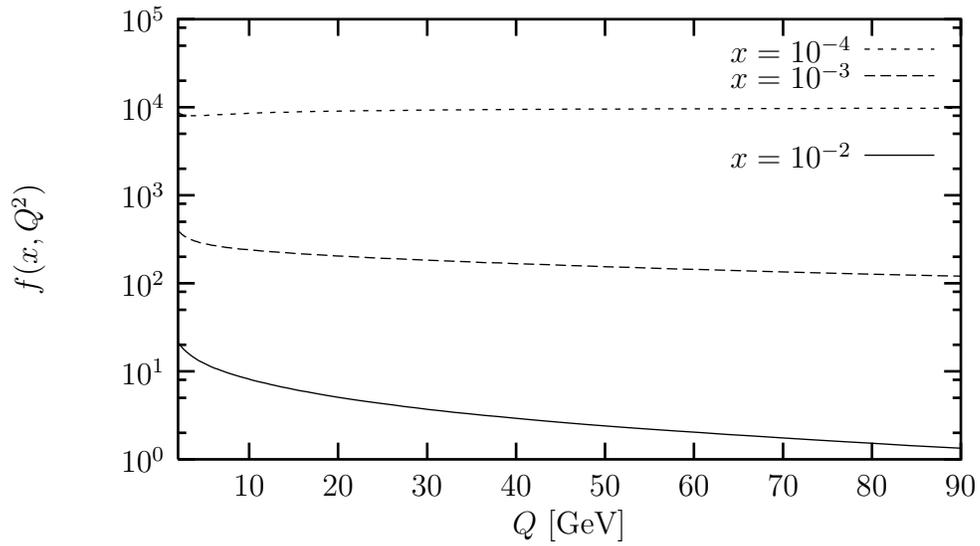

In Figs.~\ref{fig10} and \ref{fig11}, we show the resulting structure
function $f(x,Q^2)$ plotted against $x$ for different values of $Q^2$.
Fig.~\ref{fig10} refers to fixed coupling and Fig.~\ref{fig11} to
running coupling. We have also displayed a curve indicating the
$x^{-1/2}$ dependence that one would expect to obtain from the
na\"{\i}ve counting of amplitudes in which spin-$\frac{1}{2}$
particles are exchanged in the $t$-channel. The enhancement at low-$x$
is clearly seen and at sufficiently low-$x$ the structure function
behaves like
$$
f(x,Q^2) \sim x^{-1/2+\sqrt{2\tilde{\alpha}_s(Q^2)}},
$$
for the case of fixed coupling, and for running coupling we have an even
 further enhancement of the low-$x$ rise which  asymtotically
behaves like
$$
f(x,Q^2) \sim x^{-1/2+\sqrt{2\tilde{\alpha}_{\max}}}.
$$

On the other hand, we see from Figs. \ref{fig12} and \ref{fig13} that
there is much less $Q^2$ dependence in the rise at low-$x$ when the
running of the coupling is taken into account.  This is consistent
with recent results \cite{thorne} obtained in the case of the Pomeron
with running coupling. This can be attributed to the fact that for the
$Q^2$ dependence of the structure functions at low-$x$, the effect of
running the coupling is to sample values of $k^2$ which are larger
than $Q^2$ for which the running coupling is smaller than the fixed
coupling value, whereas for the $Q^2$ independent part one is also
probing the infrared region where the running coupling is enhanced.

Finally, we consider possible non-perturbative effects. One important
consequence of non-perturbative QCD is that quarks acquire a
constituent mass. As soft quarks are exchanged in the $t$-channel in
the Reggeon, it is quite likely that this will be the most significant
contribution of non-perturbative QCD to the behaviour of the Reggeon.
We also choose to give the soft gluons an effective mass, and so to
simulate the effects of non-perturbative QCD we assigning a mass, $m$,
to the soft quarks and gluons.

The kernel (\ref{boo}) then becomes
\begin{multline}
  \label{eq:ap}
  \mathcal{K}_m(s ,k,k')\otimes f(s ,k')= \int_0^1\frac{dz}{z} \Biggl[
  \tilde{\alpha}_s(k^2)\frac{ z^2}{z^2+M^2}
  \left(\frac{1+z^2+M^2}{\sqrt{((1+z)^2+M^2)((1-z)^2+M^2)}}
    -1\right)\\
  \times\left(f(s,kz)-f(s,k)\right) 
  + \tilde{\alpha}_s(k^2/z^2) \frac{ 1/z^2}{1/z^2+M^2} \\
  \times\left(\frac{1+1/z^2+M^2}{\sqrt{((1+1/z)^2+M^2)((1-1/z)^2+M^2)}}-1
  \right)\left(f(s,k/z)-f(s,k)\right)\Biggr]
  \\
  +\int_{1/s}^1\frac{dz}{z} \left[ \tilde{\alpha}_s(k^2)\frac{
      z^2}{z^2+M^2}f(s z,kz)
    +\tilde{\alpha}_s(k^2/z^2)\frac{1/z^2}{1/z^2+M^2}f(s
    z,k/z)\right],
\end{multline}
in which we have defined $M^2=m^2/k^2$.

The expression for $\omega_0(k)$ then becomes
\begin{equation}
  \omega_0(k^2)=\frac{1}{2}\int\frac{d{k'}^2}{{k'}^2+m^2}
  \tilde\alpha_s(\max(k^2,{k'}^2))
  \left[\min\left(\frac{k}{k'},\frac{k'}{k}\right)\right]^{\omega_0(k^2)}.
\end{equation}
and it is this expression for $\omega_0(k^2)$ which is used
 to define $g(s,k)$ in (\ref{gdefined}).

\begin{figure}[htb]
  \begin{center}
    \leavevmode
\setlength{\unitlength}{0.1bp}
\begin{picture}(3600,2160)(0,0)
\put(3137,1497){\makebox(0,0)[r]{$Q^2=100$}}
\put(3137,1647){\makebox(0,0)[r]{$Q^2=9$}}
\put(3137,1797){\makebox(0,0)[r]{$Q^2=1$}}
\put(3137,1947){\makebox(0,0)[r]{$Q^2=0.25$}}
\put(2000,150){\makebox(0,0){propagator mass $m$ [GeV]}}
\put(100,1230){%
\makebox(0,0)[b]{\shortstack{$\omega_0(Q^2)$}}%
}
\put(3550,300){\makebox(0,0){2}}
\put(2775,300){\makebox(0,0){1.5}}
\put(2000,300){\makebox(0,0){1}}
\put(1225,300){\makebox(0,0){0.5}}
\put(450,300){\makebox(0,0){0}}
\put(400,2060){\makebox(0,0)[r]{0.8}}
\put(400,1783){\makebox(0,0)[r]{0.7}}
\put(400,1507){\makebox(0,0)[r]{0.6}}
\put(400,1230){\makebox(0,0)[r]{0.5}}
\put(400,953){\makebox(0,0)[r]{0.4}}
\put(400,677){\makebox(0,0)[r]{0.3}}
\put(400,400){\makebox(0,0)[r]{0.2}}
\end{picture}

    \caption{The variation of the leading eigenvalue with propagator
    mass, shown for four different values of $Q^2$ (all given in GeV$^2$)}
    \label{fig:omega_mass}
  \end{center}
\end{figure}
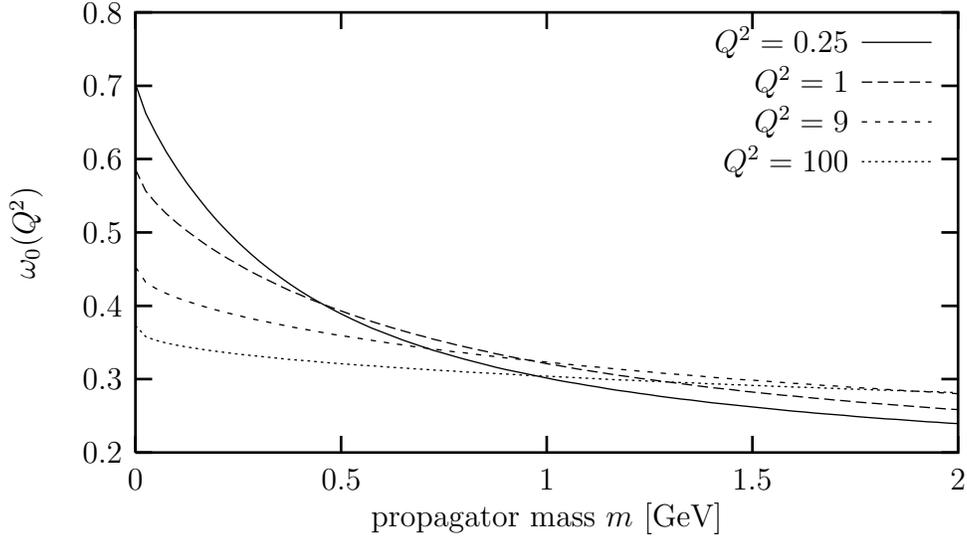

The introduction of the mass term acts to reduce the value of
$\omega_0$, and a plot of this effect is given in
Fig.~\ref{fig:omega_mass}. Note that the smaller $Q^2$ values
receive a more marked decrease in the leading eigenvalue, which further
acts to limit growth in the infra-red region.

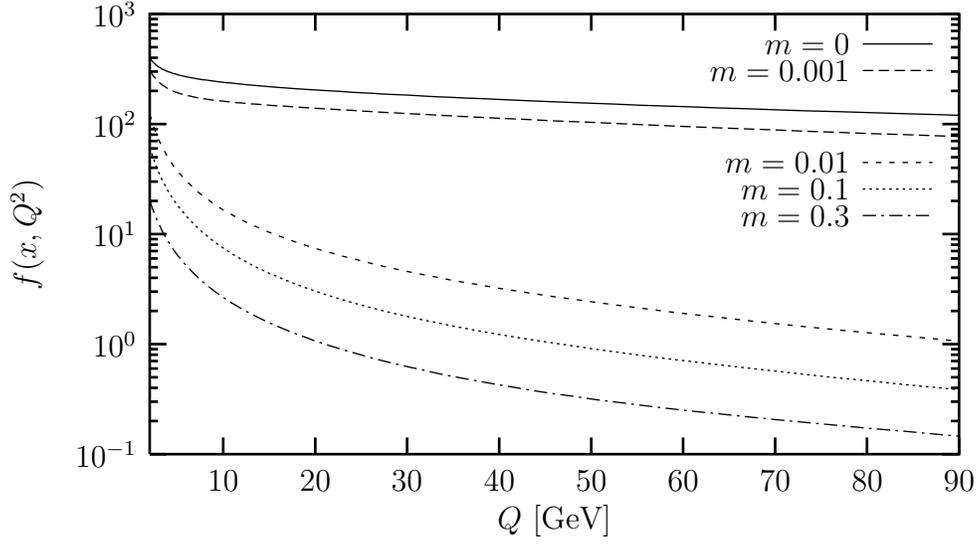
\begin{figure}[p]
  \begin{center}
    \leavevmode
\setlength{\unitlength}{0.1bp}
\begin{picture}(3600,2160)(0,0)
\put(3137,1300){\makebox(0,0)[r]{$m=0.3$}}
\put(3137,1400){\makebox(0,0)[r]{$m=0.1$}}
\put(3137,1500){\makebox(0,0)[r]{$m=0.01$}}
\put(3137,1847){\makebox(0,0)[r]{$m=0.001$}}
\put(3137,1947){\makebox(0,0)[r]{$m=0$}}
\put(2025,150){\makebox(0,0){$Q$ [GeV]}}
\put(100,1230){%
\makebox(0,0)[b]{\shortstack{$f(x,Q^2)$}}%
}
\put(3550,300){\makebox(0,0){90}}
\put(3203,300){\makebox(0,0){80}}
\put(2857,300){\makebox(0,0){70}}
\put(2510,300){\makebox(0,0){60}}
\put(2164,300){\makebox(0,0){50}}
\put(1817,300){\makebox(0,0){40}}
\put(1470,300){\makebox(0,0){30}}
\put(1124,300){\makebox(0,0){20}}
\put(777,300){\makebox(0,0){10}}
\put(450,2060){\makebox(0,0)[r]{$10^{3}$}}
\put(450,1645){\makebox(0,0)[r]{$10^{2}$}}
\put(450,1230){\makebox(0,0)[r]{$10^{1}$}}
\put(450,815){\makebox(0,0)[r]{$10^{0}$}}
\put(450,400){\makebox(0,0)[r]{$10^{-1}$}}
\end{picture}
    \caption{The effect of non-zero propagator masses is shown at
    $x=10^{-3}$. Increasing the mass (all in GeV) leads to a reduction
    at large $Q$ due to the decrease in the leading eigenvalue $\omega_0$}
    \label{fig:run_mass_fvQ}
  \end{center}
\end{figure}
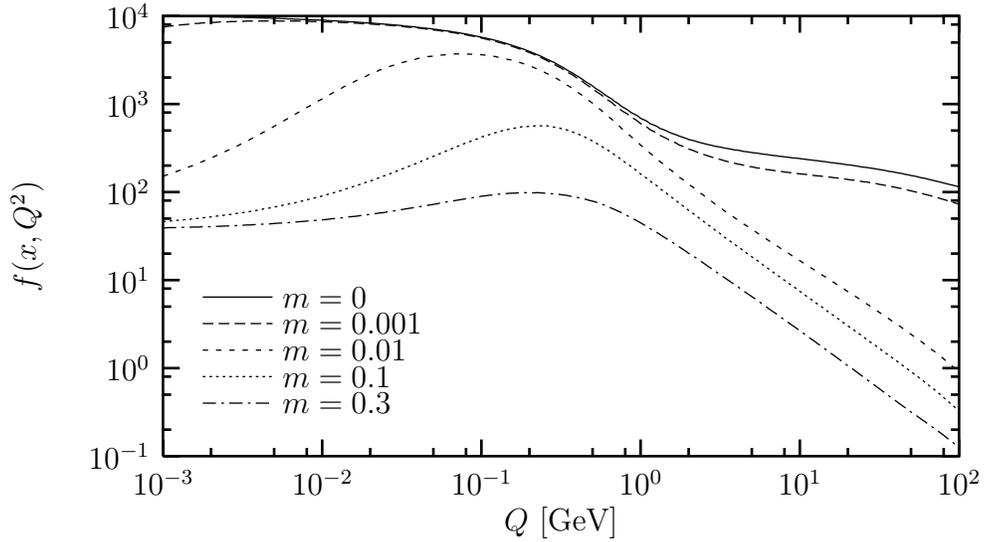
\begin{figure}[p]
  \begin{center}
    \leavevmode
\setlength{\unitlength}{0.1bp}
\begin{picture}(3600,2160)(0,0)
\put(1000,600){\makebox(0,0)[l]{$m=0.3$}}
\put(1000,700){\makebox(0,0)[l]{$m=0.1$}}
\put(1000,800){\makebox(0,0)[l]{$m=0.01$}}
\put(1000,900){\makebox(0,0)[l]{$m=0.001$}}
\put(1000,1000){\makebox(0,0)[l]{$m=0$}}
\put(2050,150){\makebox(0,0){$Q$ [GeV]}}
\put(100,1230){%
\makebox(0,0)[b]{\shortstack{$f(x,Q^2)$}}%
}
\put(3550,300){\makebox(0,0){$10^{2}$}}
\put(2950,300){\makebox(0,0){$10^{1}$}}
\put(2350,300){\makebox(0,0){$10^{0}$}}
\put(1750,300){\makebox(0,0){$10^{-1}$}}
\put(1150,300){\makebox(0,0){$10^{-2}$}}
\put(550,300){\makebox(0,0){$10^{-3}$}}
\put(500,2060){\makebox(0,0)[r]{$10^{4}$}}
\put(500,1728){\makebox(0,0)[r]{$10^{3}$}}
\put(500,1396){\makebox(0,0)[r]{$10^{2}$}}
\put(500,1064){\makebox(0,0)[r]{$10^{1}$}}
\put(500,732){\makebox(0,0)[r]{$10^{0}$}}
\put(500,400){\makebox(0,0)[r]{$10^{-1}$}}
\end{picture}
    \caption{The same data as in figure~\ref{fig:run_mass_fvQ} is
      shown here with a logarithmic x-axis to demonstrate the
      infra-red effects of the propagator mass}
    \label{fig:run_mass_fvQ_log}
  \end{center}
\end{figure}
We show the effect of the massive propagators in
Fig.~\ref{fig:run_mass_fvQ}. Here we have plotted against $Q$ for five
values of $m$, including $m=0$. The effect is quite strong, even at
larger values of $Q$, with the reduction in $f$ being more pronounced
than might be expected of a measure originally designed to affect just
the infra-red region. The reason behind this has of course just been
detailed: increasing $m$ reduces $\omega_0$ so that by the time we
have evolved the structure function to low-$x$ (the graph is at
$x=10^{-3}$) it is substantially reduced across its whole domain due
to the weaker leading behaviour. If we were to consider the same
function at larger $x$ the differences between the graphs for
differing $m$ would not be so pronounced for large $Q^2$.

In Fig.~\ref{fig:run_mass_fvQ_log} we have shown exactly the same
data as in the previous graph, but with the $Q$ scale now logarithmic
in order to illustrate the small $Q$ region more clearly. As expected,
the introduction of the mass term results in a reduction at low $Q$,
but whereas the larger $Q$ reduction only becomes apparent as we
evolve to low-$x$, the growth in the $Q\sim m$ region is immediately
regulated.

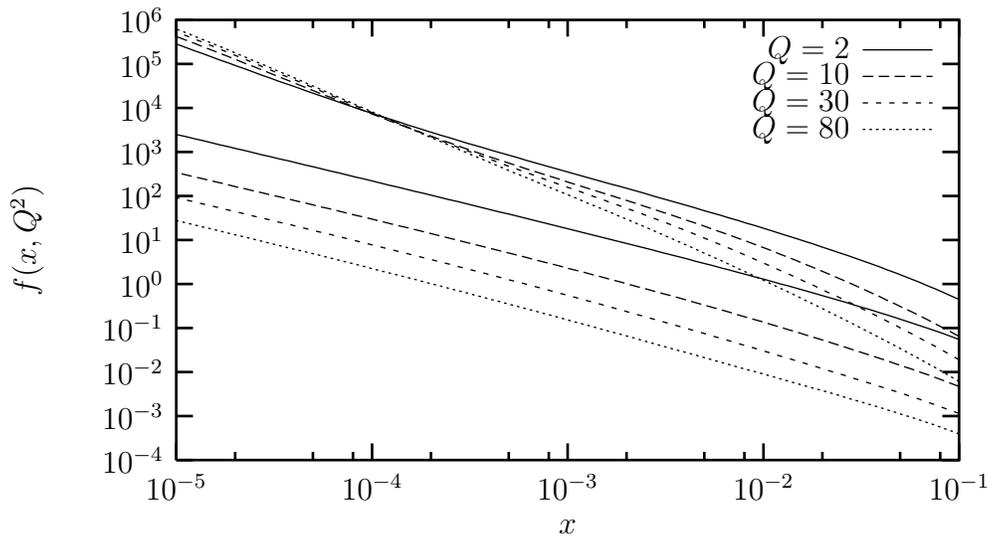
\begin{figure}[htbp]
  \begin{center}
    \leavevmode
\setlength{\unitlength}{0.1bp}
\begin{picture}(3600,2160)(0,0)
\put(3137,1647){\makebox(0,0)[r]{$Q=80$}}
\put(3137,1747){\makebox(0,0)[r]{$Q=30$}}
\put(3137,1847){\makebox(0,0)[r]{$Q=10$}}
\put(3137,1947){\makebox(0,0)[r]{$Q=2$}}
\put(2075,150){\makebox(0,0){$x$}}
\put(100,1230){%
\makebox(0,0)[b]{\shortstack{$f(x,Q^2)$}}%
}
\put(3550,300){\makebox(0,0){$10^{-1}$}} 
\put(2813,300){\makebox(0,0){$10^{-2}$}}
\put(2075,300){\makebox(0,0){$10^{-3}$}}
\put(1338,300){\makebox(0,0){$10^{-4}$}}
\put(600,300){\makebox(0,0){$10^{-5}$}}
\put(550,2060){\makebox(0,0)[r]{$10^{6}$}}
\put(550,1894){\makebox(0,0)[r]{$10^{5}$}}
\put(550,1728){\makebox(0,0)[r]{$10^{4}$}}
\put(550,1562){\makebox(0,0)[r]{$10^{3}$}}
\put(550,1396){\makebox(0,0)[r]{$10^{2}$}}
\put(550,1230){\makebox(0,0)[r]{$10^{1}$}}
\put(550,1064){\makebox(0,0)[r]{$10^{0}$}}
\put(550,898){\makebox(0,0)[r]{$10^{-1}$}}
\put(550,732){\makebox(0,0)[r]{$10^{-2}$}}
\put(550,566){\makebox(0,0)[r]{$10^{-3}$}}
\put(550,400){\makebox(0,0)[r]{$10^{-4}$}}
\end{picture}
    \caption{The upper set of four lines show zero mass results to
      provide comparison with the lower set which have $m=0.3$ GeV. As
      expected the rise at low $x$ has been slowed by the non-zero
      mass}
    \label{fig:mass_fvx}
  \end{center}
\end{figure}

In Fig.~\ref{fig:mass_fvx} we have again plotted the zero
mass/running coupling graphs for four $Q$ values, but we have also
included on the same plot the case where we take $m=0.3$ GeV. Thus the
upper set of lines are the same as in figure~\ref{fig11}, and
the lower set show the relative effect of the propagator's mass. As
would be expected from our previous arguments the low-$x$ rise in $f$
is reduced.

\section{Summary}

We have reviewed the derivation of the reggeization of the quark. This
reggeized quark is used to construct the integro-differential equation
for the amplitude of a process in which the quantum numbers of the
Reggeon are exchanged. We have proposed a method for the numerical
solution of this equation which is consistent with the leading large
$s$ behaviour. This enables one to make modifications to the kernel in
order to account for a running coupling and for simulations of
non-perturbative effects.

The experimental quantities which can be used for probing the QCD
simulated Reggeon are flavour non-singlet quantities at low-$x$. In
all probability the reach of current experiments in deep-inelastic
scattering are not yet sufficient provide us with such a probe, but it
may well become possible in the near future.

We find from our numerical studies that the introduction of the
running of the coupling enhances the already sharp rise expected in
the structure functions as $x \to 0$. On the other hand the $Q^2$
dependence is considerably moderated when a running coupling is
introduced. The effects of QCD beyond perturbation theory have been
estimated by assigning a mass to the soft particles exchanged. We find
that this has a dramatic effect in reducing the rise at low-$x$, even
at values of $Q^2$ which are considerably larger than the values of
the constituent mass inserted. All of this suggests that it will be
even harder to isolate the kinematic region where we might expect the
results of perturbative QCD to dominate, and although a rise at
low-$x$ is still expected, it may well not be as dramatic as initially
expected.

\end{document}